\documentclass[fleqn,usenatbib]{mnras}

\usepackage{newtxtext,newtxmath}

\usepackage[T1]{fontenc}

\DeclareRobustCommand{\VAN}[3]{#2}
\let\VANthebibliography\thebibliography
\def\thebibliography{\DeclareRobustCommand{\VAN}[3]{##3}\VANthebibliography}

\usepackage{graphicx}	
\usepackage{amsmath}	

\title[M14 horizontal branch and beyond]{A new photometric study of M14 (NGC 6402): An interpretation of the Horizontal Branch and beyond
$\thanks{Based on observations made with the 0.84m telescope at San Pedro M\'artir Observatory, Baja California
(Mexico) and the 2.0m Chandra telescope at the Indian Astrophysical
Observatory, Hanle (India).}$}

\author[Yepez et al.]{
M. A. Yepez,$^{1}$\thanks{E-mail: myepez@astro.unam.mx} A. Arellano Ferro,$^{1}$
D. Deras$^{1}$, I. Bustos Fierro$^{2}$, \and S. Muneer$^{3}$, K.-P. Schr\"oder$^{4}$
\\
$^{1}$Instituto de Astronom\'ia, Universidad Nacional Aut\'onoma de M\'exico, Ciudad de M\'exico, CP 04510, M\'exico.\\
$^{2}${Universidad Nacional de C\'ordoba. Observatorio
Astron\'omico. C\'ordoba,  Argentina.}\\
$^{3}${Indian Institute of Astrophysics. Bangalore, India.}\\
$^{4}$Departamento de Astronom\'ia, Universidad de Guanajuato, M\'exico.\\
}

\date{Accepted -- 27. Received --; in original form --}

\pubyear{2020}

\begin{document}
\label{firstpage}
\pagerange{\pageref{firstpage}--\pageref{lastpage}}
\maketitle

\begin{abstract}
We present a CCD \emph{VI} photometric study of the globular cluster M14. Particular attention is given to the variable stars. This allowed new classifications and cluster membership considerations. New variables are reported; 3 RRc, 18 SR and 1 SX Phe. The Fourier decomposition of RR Lyrae light curves lead to the mean cluster metallicity of [Fe/H]$_{\rm ZW}= -1.3 \pm 0.2$. Several independent methods yield a mean distance of $9.36 \pm 0.16$ kpc.  A Colour-Magnitude diagram outlined by the cluster members enabled a matching with theoretical predictions of isochrones and zero-age horizontal branches, whose fitting to the observations is in good agreement with the above distance and metallicity. The Oosterhoff type of M14 is confirmed as Oo-int, and the pulsating mode distribution of RR Lyrae stars on the HB shows that the bimodal region of the instability strip is shared by RRab and RRc stars. By modelling the mass loss at the RGB after the He flash events, we were able to represent the blue tail of the HB, using a core mass of 0.48 $M_{\odot}$ and total masses of 0.52-0.55 $M_{\odot}$. A progenitor star on the MS of 0.84 $M_{\odot}$ reaches the HB in about 12.5 Gyrs, consonant with previous age determinations of the cluster. Type II Cepheids of M14 may be interpreted as products of post-HB evolution, driven by the complex processes involving the burning of the very thin low mass hydrogen and helium shells of these stars and their minuscule envelopes. No evidences were found in favor of M14 being of extragalactic origin.
\end{abstract}




\section{Introduction}
The variable stars in globular clusters have proven to be very useful as indicators of quantities relevant to the cluster properties, such as distance, reddening and metallicity, but also as indicators of individual stellar physical parameters, such as effective temperature, luminosity, mass and radius. Numerous works can be found in the literature where efforts in these directions have been invested. Probably the best source in the literature to trace the whole historical developments per cluster and for a family of 152 globular clusters is the Catalogue of Variable Stars in Globular Clusters (CVSGC), continuously updated by Prof. Christine Clement \citep{cle01}. Clusters rich in variable stars are of particular interest if structural and evolutionary properties of the several families of variables are in the scope or if their implications of some of the system properties are pursued (e.g. \cite{aaf20}; \cite{Yepez20} and references therein).

The Globular Cluster M14 (NGC 6402) ($\alpha$ = 17:37:36.1, $\delta$ = -03:14:45.3, J2000) is a very rich cluster in variable stars, particularly of the RR Lyrae type. In the CVSGC, there are 173 variables listed, although for a few of them their variability has not been confirmed. The cluster is relatively close to the Sun, with a recorded distance of 9.3 kpc, and of intermediate metallicity [Fe/H]=$-1.28$  \citep{Harris1996}, a convenient circumstance to study the variable stars light curves. In spite of its short distance, the cluster seems to be subject to a substantial reddening, $E(B-V) \sim 0.57-0.62$, although its precise value is rather controversial and some evidence of differential reddening has been reported \citep{CP2013} (hereinafter CP13).\\ 

The variable star population of M14 has been studied for more than 80 years, beginning with \citet{Sawyer1938} who identified the first 72 variables but gave periods for only three stars. \citet{Sawyer1966} found another 4 variables and reported periods for 20 variables. \citet{Wehlau1972} gave coordinates and amplitudes for 12 more new variables. \citet{Wehlau1974} found 5 more variables by blinking $B$ filter plates. \citet{Wehlau1994}, using photographic $B$ photometry with a time-base of 68 years, catalogued 68 of these variables and offered periods and period change rates for numerous RR Lyrae stars. \citet{Conroy2012} found another 71 variables. The most recent work on the variables of M14 was carried out by \citet{CP2018} (hereinafter CP18), where they reclassified the variable status of some stars and found 8 new ones. The final count of variables prior to the present paper is 133 \citep{cle01} distributed by types as: 56 RRab, 54 RRc, 6 type II Cepheids (CW), 3 Eclipsing Binaries, and 14 long-period variables. No SX Phe were known at that time in M14.\\

A significant peculiarity of M14 is that its HB has a sound and well developed blue morphology, in spite of its metallicity being [Fe/H]=$-1.28$ \citep{Harris1996}. This makes a good case of the need of a second parameter, other than metallicity, controlling the structural morphology of the HB, since extreme blue morphologies are a characteristic of metal weak OoII type clusters.
These properties seem to place the cluster in the Oosterhoff gap, a region voided of galactic clusters but
otherwise occupied by clusters of 
extragalactic nature of intermediate Oosterhoff type or Oo-int \citep{Catelan2009}. Its classification as Oo-int (CP18) and the analysis of its dynamical properties \citep{Gao2007}, trigger the possibility of the cluster being of extragalactic origin.

However, more recently, making use of the kinematic information provided by $Gaia$-DR2, \citet{Massari2019} studied the possible origin of the Galactic globular cluster, and identified M14 with a low-energy system, not yet associated to known debris caused by extragalactic mergers. Later in this paper we shall reconsider and discuss the HB structural parameter in the light of a dynamical selection of cluster star members.

In this paper, we perform \emph {VI} photometry for about 9750 stars in the field of the cluster and present the light curves of 121 previously known variables and of 22 newly discovered. We made use of the difference image analysis (DIA) technique and the DanDIA pipeline \citep{Bramich2008,Bramich2013} to recover accurate photometry of point sources in our images. We aim to extract physical parameters of individual stars and reliable estimation of the mean reddening, metallicity and distance to the cluster. The resulting Colour-Magnitude diagram (CMD) and the structure of the Horizontal Branch (HB) are discussed in detail and are confronted with evolutionary predictions. In the process, we explored the properties of some peculiar stars, such as the double mode nature of several variables, corrected the coordinates and identifications of some stars using data of the $Gaia$-eDR3 and performed a membership analysis of the stars measured in the FoV of our images. 
The paper is organised as follows: In $\S$ 2 the observations and reduction process are described. In $\S$ 3 An estimation of the cluster reddening is offered. $\S$ 4 The several families of variable stars are discussed and the strategies and results of search of new variables are presented. $\S$ 5 The Fourier RR Lyrae light curve decomposition and the resulting physical parameters are reported. $\S$ 6 The Oosterhoff type of the cluster is reassessed. In $\S$ 7 the several approaches to the cluster distance determination are discussed. $\S$ 8 contains a brief description of the membership analysis using $Gaia$-eDR3. Finally in $\S$ 9 we present our conclusions on the structural appearance of the CMD, particularly the HB, and the mode distribution in the instability strip. The paper is closed with the calculation of post He flash evolution models and Zero Age Horizontal Branch (ZAHB) loci for several He core and envelope masses. Modelling from the extreme blue tail of the HB towards the type II cepheids region is also discussed. 

\section{Observations and reductions}
The CCD $VI$ observations of M14 employed in this work were obtained at two sites. The first set of observations were performed with the 0.84-m telescope at the San Pedro M\'artir observatory, in Baja California, M\'exico in two different seasons: between June 2018 and August 2018 (hereafter SPM18) the detector used was a Spectral Instruments CCD of 1024$\times$1024 pixels with a scale of 0.444 arcsec/pix, resulting in a field of view (FoV) of approximately 7.57$\times$7.57~arcmin$^2$, and between May 2019 and July 2019 (hereafter SPM19) when the detector was a Marconi5 CCD of 1024 $\times$1032 pixels with a scale of 0.493 arcsec/pixel for a FoV of $8.41\times8.48$ arcmin$^2$. The second set of observations made between August 2020 and October 2020 (hereafter Han20) come from 2.0-m telescope at the Indian Astronomical Observatory (IAO), Hanle, India. The detector used was a Thompson grade 0 ED2V 4482-0-E93 CCD of 2048$\times$2048 pixels with a scale of 0.296 arcsec/pix, translating into a FoV of approximately 10.1$\times$10.1~arcmin$^2$. A total of 874 and 1047 images were obtain in $V$ and $I$ filters, respectively. The log of the observations, exposure times and seeing conditions are summarised in Table \ref{tab:obs}.\\

\begin{table}
\caption{The observations of M14 at SPM (2018-2019) and IAO (2020)$^*$}
\centering
\begin{tabular}{lcccccc}
\hline
Date  & HJD& $N_{V}$ & $t_{V}$ & $N_{I}$ &$t_{I}$&seeing \\
      &245 0000.+   &       & sec    &         &sec& (")\\
\hline
18-06-09 &8279.25 &66 &50 &82 & 30  & 2.0\\
18-06-17 &8287.25 &91 &60 &99 & 40 & 1.9 \\
18-07-15 &8315.25 &93 &60 &86 & 40 & 1.8 \\
18-07-24 &8323.25   & 6 & 60 & 6&40 & 1.7\\
18-08-11 & 8342.25&31 &60  & 28  & 40  & 2.0\\
18-08-12 & 8343.25& 18 &60  & 17  &40   &1.9\\
18-08-13 & 8344.25&21 &60   &  28 & 40 &1.7\\
18-08-14 & 8345.25&33 & 60  &  40 & 40  &1.7\\
19-05-26 &8630.25& 29 &60 &37& 40  & 2.0\\
19-05-27 &8631.25& 1 & 60& 2 & 40 & 3.4\\
19-05-28 &8632.25& 61&60& 82 & 40 &1.9\\
19-05-29 &8633.25&41 &60 & 56 & 40 &2.4\\
19-05-30 &8634.25&59 &60& 80 & 40 &2.2\\
19-06-25 &8660.25& 50 & 60 & 60 & 40 &1.9\\ 
19-06-26 &8661.25& 65 & 60 & 87 & 40 & 1.5\\
19-06-27 &8662.25& 39 & 60 & 60 & 40 &1.8\\
19-06-28 &8663.25& 42 & 60 & 53 & 40 &2.0\\
19-07-01 &8666.25& 44 & 60 & 67 & 40 &1.9\\
20-08-14 &9076.25&26 &10,100  &26 &40,400  &2.0\\
20-08-16 &9078.25&30 &10,100  & 32 &40,400  &1.5\\
20-09-17 &9110.25&19&10,40  &12  &40,180  &2.2\\
20-10-11 &9134.25& 9 &10  &7 &40  &2.8\\
\hline
Total:  & & 874&    & 1047 &    & \\
\hline
\end{tabular}
\raggedright
\center{\quad $*$Columns: HJD at the beginning of the night. $N_{V}$ and $N_{I}$ give the number of images taken with the $V$ and $I$
filters respectively. $\MakeLowercase{t}_{V}$ and $\MakeLowercase{t}_{I}$
provide the exposure time. In the last column the average nightly average seeing is indicated.}
\label{tab:obs}
\end{table}

The image reduction was performed using the Difference Image Analysis (DIA) through the pipeline DanDIA \citep{Bramich2008,Bramich2013}. This approach has been repeatedly used and described in detail in previous publications, and the interested reader is referred for instance to the paper by \citet{Bramich2011}.\\
 
\subsection{Transformation to the photometric standard system}

We transformed our instrumental \emph{vi} photometry into the Johnson-Kron-Cousins (JKC) standard \emph{VI} system using the standard stars in the FoV of M14 included in the catalogue of Photometric Standard Fields \citep{Stetson2000}.

\begin{figure}
\begin{center}
\includegraphics[width=8cm]{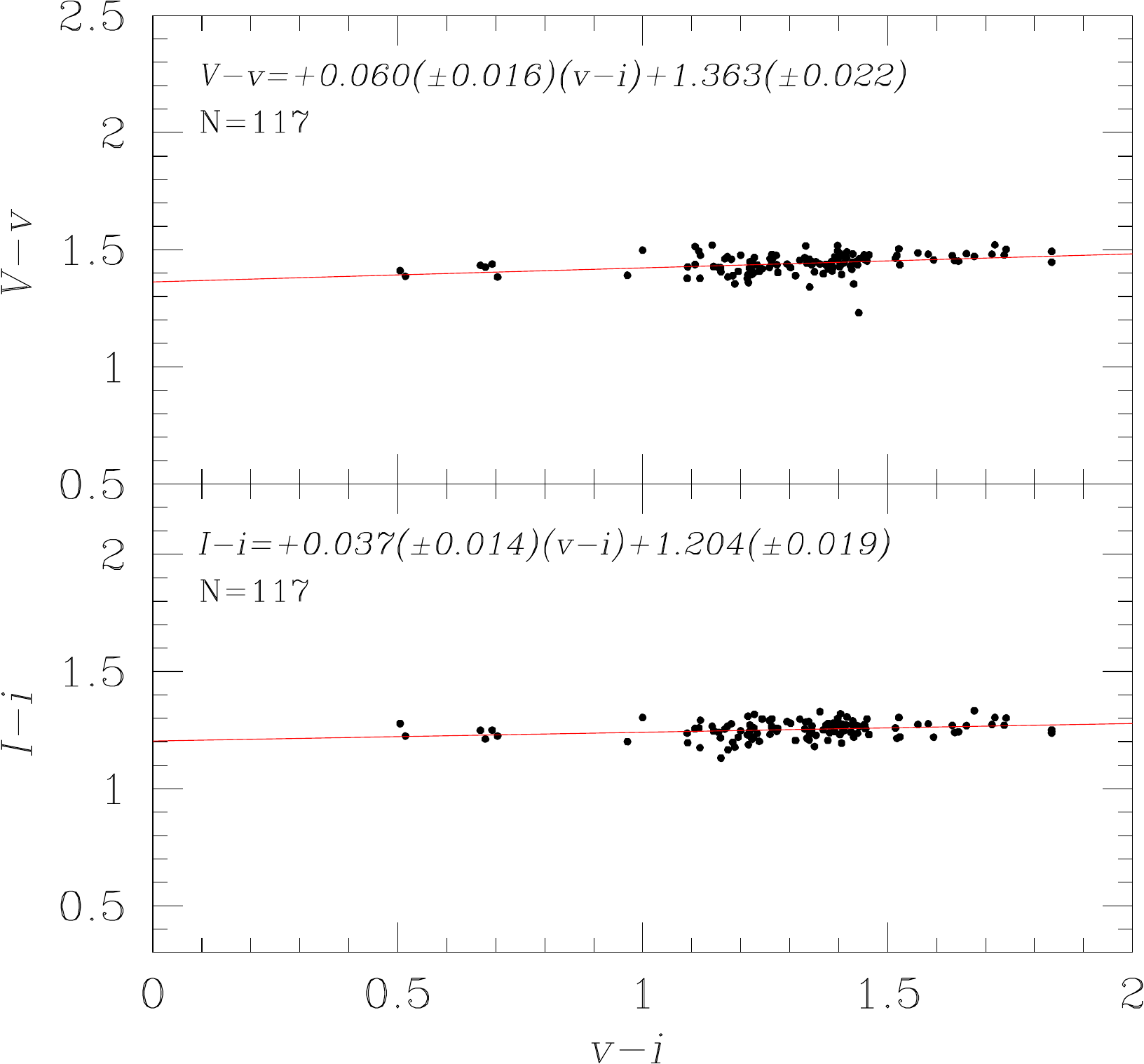}
\caption{Transformation equations between instrumental and standard photometric systems. The equations were calculated based on 117 standard stars from the collection of Stetson (2000) in the FoV of Han20 data set.}
    \label{transcolor}
\end{center}
\end{figure}

We identified 51, 46 and 117 standard stars in the SPM18, SPM19 and Han20 fields, respectively. Fig. \ref{transcolor} is an example of the transformation for the case of Han20 data. The \emph{VI} transformation equations are shown in the figure itself. A small colour term is present.\\

\section{An approach to the cluster reddening}
\label{reddening}

In spite of its relative close proximity to the Sun, being near the Galactic bulge ($l = 21.32^{\circ}$, $b = -14.81^{\circ}$) the cluster is subject to a substantial reddening,  although its value is rather controversial.  For example, \citet{Harris1996} lists a value of $E(B-V) = 0.60$, but a much lower value $\sim 0.41 \pm 0.01$  is  reported  by  the  calibration of \citet{Schlafly2011}. In a detailed investigation CP13 concluded that the cluster  has  an  average reddening of $E(B-V)  =  0.57 \pm 0.02$,  and that it is subject of differential reddening, amounting as much as $\Delta E(B-V) = 0.17$ mag. CP13 provide a differential reddening map of the field of M14, which we employed to correct the CMD as explained lated in $\S$ \ref{secCMD}.
  
We have approached the reddening problem, by taking advantage of the numerous RRab stars in the cluster and of the fact that these stars have a constant intrinsic colour $(V-I)_0$ near minimum light, between phases 0.5 and 0.8 \citep{Sturch1966}. The intrinsic value $(V-I)_0$ in this range of phases was calibrated  by \citet{Guldenschuh2005} who found a value of  $\overline{(V-I)_{o,min}}$ = 0.58 $\pm$ 0.02 mag. Hence, by measuring the average $\overline{(V-I)_{o,0.5-0.8}}$ one can estimate $E(V-I) =\overline{(V-I)_{o,min}} - (V-I)_{o,0.5-0.8}$ for each individual RRab star with a properly covered colour curve at these phases. Then we calculated $E(B-V)$ via the ratio $E(V-I)/E(B-V) = 1.259$.

We included in this calculation only those RRab stars with the phase interval 0.5-0.8 sufficiently well covered and without evident signs of amplitude modulations. A total of 28 stars were considered for an average $E(B-V)= 0.618 \pm 0.048$. This value, although within the uncertainties is in good concordance with the previous estimations given above, it shall prove however to be a bit too large when it comes to the distribution of stars in the HB and instability strip and isochrones positions in the CMD and the H-R Diagram (HRD). An average overall value of $E(B-V)= 0.59$ will produce a better agreement between the data and some theoretical predictions, as will be shown in $ \S$ \ref{secCMD}. 

\section{The variable star population of M14}
 
\subsection{Sample of known variables included in this work}

The population of variable stars in M14 is vast. The CVSGC  lists 172 variable stars and one nova not numbered, 39 of which have been found to be constant or repeated from previous works \citep{Conroy2012,CP2018}. Variables V25, V27, V28, V36, V51, V163 and V164 are not considered in this work as they are out of the FoV of our images.\\

In order to correctly identify all the variables, first we calibrated our reference image astrometrically, and loaded equatorial coordinates from the CVSGC and $Gaia$-eDR3, along with the X-Y coordinates of our light curves to establish the correspondence.
By blinking the residual images, we found that for stars V38, V121, V127, V128, and V144, our DanDIA photometry was unable to recover the proper variable due to the presence of blended neighbours, hence we shall not consider these variables any further. This methodology allowed us to refine the coordinates of all variables known in M14, including the variables out of the FoV of our images. The $Gaia$ equatorial coordinates corresponding to the variable stars are listed in Table \ref{tab:variables}.

\begin{table*}
\small
\scriptsize
\begin{center}
\caption{Mean magnitudes, amplitudes, periods and equatorial coordinates of variable stars in M14 in the FV of our images. A cluster membership flag is also included in the last column.}
\label{tab:variables}
\begin{tabular}{llccccccccc}
\hline
ID & Variable & $<V>^1$ & $<I>^1$ & $A_V$ & $A_I$ & $P^2$ (days) &  HJD$_{\rm max}$ &  RA  &  Dec. & mem$^3$\\
  &  Type  & (mag) & (mag) & (mag) & (mag) & this work &  (2450000+)  & (J2000.0) & (J2000.0)&\\
\hline
V1       & CW   & 14.466 & 12.780 & 1.197 & 0.925 & 19.7411  & 8633.9408 & 17:37:37.38 & -03:13:59.5 & m  \\
V2       & CW   & 15.700 & 14.323 & 0.806 & 0.608 & 2.7949   & 8343.7760 & 17:37:28.61 & -03:16:45.2 & m  \\
V3       & RRd  & 16.637 & 15.484 & 0.665 & 0.535 & 0.522380 & 8665.7971 & 17:37:36.07 & -03:16:14.4 & ?  \\
         &      &        &        &       &       & 0.403379 &           &             &             &    \\
V4       & RRab & 17.205 & 15.926 & 0.967 & 0.700 & 0.651313 & 8278.8355 & 17:37:47.01 & -03:13:31.9 & m  \\
V5       & RRab & 17.276 & 16.058 & 1.251 & 0.790 & 0.548793 & 8278.9163 & 17:37:27.19 & -03:13:17.8 & m  \\
V6       & SR   & 14.306 & 11.779 & 0.181 & 0.110 & 69.8565  & 8278.8852 & 17:37:38.57 & -03:16:02.5 & m  \\
V7       & CW   & 14.817 & 13.163 & 0.830 & 0.723 & 13.5897  & 8278.8812 & 17:37:40.42 & -03:16:20.5 & m  \\
V8       & RRab & 17.222 & 15.910 & 0.600 & 0.358 & 0.686074 & 8286.8034 & 17:37:42.64 & -03:14:09.5 & m  \\
V9       & RRab & 17.281 & 16.016 & 1.024 & 0.812 & 0.538795 & 8631.9052 & 17:37:46.38 & -03:15:23.7 & m  \\
V10      & RRab & 17.083 & 15.875 & 1.050 & 0.892 & 0.585940 & 8342.7834 & 17:37:33.01 & -03:18:09.6 & m  \\
V11      & RRab & 16.755 & 15.611 & 1.106 & 0.712 & 0.604423 & 8342.7684 & 17:37:49.35 & -03:18:26.1 & f  \\
V12      & RRab & 17.247 & 16.062 & 1.225 & 0.810 & 0.503965 & 8278.8461 & 17:37:51.31 & -03:17:40.6 & m  \\
V13      & RRab & 17.040 & 15.752 & 1.160 & 0.682 & 0.535237 & 8286.8820 & 17:37:34.42 & -03:16:43.9 & m  \\
V14      & RRab & 17.468 & 16.253 & 1.464 & 0.975 & 0.472223 & 9076.1986 & 17:37:39.80 & -03:14:48.1 & m  \\
V15      & RRab & 17.250 & 16.023 & 1.024 & 0.968 & 0.557746 & 8659.7608 & 17:37:27.30 & -03:12:19.2 & m  \\
V16      & RRab & 16.873 & 15.592 & 0.890 & 0.606 & 0.600622 & 8314.7966 & 17:37:31.02 & -03:15:21.8 & m  \\
V17      & CW   & 14.642 & 13.162 & 0.543 & 0.494 & 12.0758  & 9076.1057 & 17:37:21.08 & -03:12:45.1 & m  \\
V18      & RRab & 17.376 & 16.162 & 1.453 & 0.975 & 0.478971 & 8278.8805 & 17:37:40.45 & -03:15:07.1 & m  \\
V19      & RRab & 17.216 & 16.003 & 1.190 & 0.964 & 0.545684 & 8286.7554 & 17:37:27.78 & -03:14:44.3 & m  \\
V20      & RRc  & 17.331 & 16.329 & 0.542 & 0.340 & 0.263532 & 8314.8344 & 17:37:26.57 & -03:13:08.6 & m  \\
V21      & RRc  & 16.978 & 15.985 & 0.486 & 0.303 & 0.318871 & 8629.8683 & 17:37:41.01 & -03:12:40.2 & m  \\
V22      & RRab & 17.183 & 15.926 & 0.909 & 0.601 & 0.655912 & 8632.7995 & 17:37:40.93 & -03:13:10.7 & m  \\
V23      & RRab & 17.066 & 16.126 & 0.946 & 0.597 & 0.552217 & 9078.2087 & 17:37:41.17 & -03:10:05.0 & m  \\
V24      & RRab & 17.271 & 16.126 & 1.275 & 0.933 & 0.519905 & 8659.8329 & 17:37:36.13 & -03:13:30.6 & m  \\
V25$^4$  & RRc  & -      & -      & -     & -     & -        & -         & 17:37:34.56 & -03:19:57.8 & m  \\
V27$^4$  & SR   & -      & -      & -     & -     & -        & -         & 17:37:08.10 & -03:12:17.5 & m  \\
V28$^4$  & E    & -      & -      & -     & -     & -        & -         & 17:37:05.02 & -03:08:37.6 & f  \\
V29      & SR   & \textit{14.282} & \textit{11.742} & - & -  &  -  & -   & 17:37:31.83 & -03:17:16.2 & m  \\
V30      & RRab & 17.174 & 15.969 & 1.221 & 0.864 & 0.534225 & 9076.1318 & 17:37:41.30 & -03:14:57.1 & m  \\
V31      & RRab & 17.208 & 15.926 & 1.046 & 0.722 & 0.619636 & 8314.7400 & 17:37:33.63 & -03:14:14.1 & m  \\
V32      & RRab & 16.993 & 15.833 & 0.877 & 0.594 & 0.655977 & 8286.8801 & 17:37:38.56 & -03:12:18.8 & m  \\
V33      & RRab & 17.290 & 16.130 & 1.130 & 0.762 & 0.479938 & 8286.8053 & 17:37:27.02 & -03:14:32.2 & m  \\
V34      & RRab & 17.323 & 15.991 & 0.891 & 0.569 & 0.606620 & 8661.7707 & 17:37:31.55 & -03:14:19.3 & m  \\
V36$^4$  & RRab & -      & --     & -     & --    & --       & -         & 17:37:50.05 & -03:20:29.3 & m  \\
V37      & RRab & 17.256 & 15.908 & 1.026 & 0.575 & 0.489034 & 8314.8514 & 17:37:36.68 & -03:14:27.1 & m  \\
V38      & RRab & -      & -      & -     & -     & -        & -         & 17:37:36.98 & -03:15:02.7 & ?  \\
V39      & RRab & 17.270 & 15.962 & 1.312 & 0.835 & 0.576010 & 8630.7462 & 17:37:39.32 & -03:14:45.8 & m  \\
V41      & RRc  & 17.235 & 16.140 & 0.500 & 0.289 & 0.259343 & 8630.7455 & 17:37:35.09 & -03:14:47.3 & m  \\
V42      & RRab & 16.926 & 15.859 & 0.821 & 0.659 & 0.631103 & 8286.809  & 17:37:38.70 & -03:14:33.3 & m  \\
V43      & RRab & 17.393 & 16.172 & 1.286 & 0.987 & 0.521738 & 8278.8698 & 17:37:40.69 & -03:14:22.9 & m  \\
V44      & RRc  & 17.120 & 16.152 & 0.511 & 0.303 & 0.289420 & 8314.8382 & 17:37:37.50 & -03:12:47.7 & m  \\
V45      & SR   & \textit{14.358} & \textit{12.023} & -  & -  &  -  & -  & 17:37:30.18 & -03:13:12.7 & m  \\
V46      & RRc  & 17.215 & 16.114 & 0.491 & 0.333 & 0.332621 & 8633.9018 & 17:37:42.31 & -03:15:49.7 & m  \\
V47      & RRab & 17.119 & 15.727 & 0.480 & 0.357 & 0.876970 & 9078.1579 & 17:37:30.28 & -03:14:18.7 & m  \\
V48      & RRab & 17.660 & 16.335 & 0.892 & 0.844 & 0.467820 & 8286.8590 & 17:37:35.84 & -03:14:04.9 & m? \\
V49      & RRab & 16.829 & 15.413 & 0.535 & 0.261 & 0.641486 & 8632.8255 & 17:37:29.86 & -03:15:04.4 & f  \\
V51$^4$  & RRc  & -      & -      & -     & -     & -        & -         & 17:37:43.37 & -03:19:50.2 & m  \\
V55      & RRc  & 17.191 & 16.111 & 0.446 & 0.290 & 0.337565 &  -        & 17:37:38.43 & -03:12:59.0 & m  \\
V56      & RRc  & 17.039 & 15.974 & 0.475 & 0.298 & 0.341131 & 9076.2584 & 17:37:31.86 & -03:17:48.7 & m  \\
V57      & RRab & 17.200 & 15.946 & 0.946 & 0.549 & 0.566891 & 8278.7659 & 17:37:45.25 & -03:16:39.4 & m  \\
V58      & RRc  & 16.934 & 15.852 & 0.421 & 0.255 & 0.417508 &  -        & 17:37:28.11 & -03:15:19.3 & m  \\
V59      & RRab & 17.444 & 16.136 & 1.035 & 0.775 & 0.555614 & 8286.7736 & 17:37:34.14 & -03:14:16.2 & m  \\
V60      & RRab & 17.370 & 16.039 & 0.981 & 0.706 & 0.578874 & 8278.7805 & 17:37:39.04 & -03:13:50.4 & m  \\
V61      & RRab & 17.000 & 15.765 & 0.843 & 0.505 & 0.569823 & 8341.7977 & 17:37:37.19 & -03:15:28.3 & m  \\
V62      & RRab & 17.288 & 15.973 & 0.420 & 0.255 & 0.638512 & 8314.8457 & 17:37:20.92 & -03:17:20.7 & m  \\
V68      & RRab & 17.47  & 16.280 & 1.500 & 0.861 & 0.507229 & 8286.7650 & 17:37:36.89 & -03:15:05.4 & m  \\
V70      & RRab & 17.224 & 16.004 & 1.122 & 0.705 & 0.604915 & 9076.2727 & 17:37:39.11 & -03:15:07.7 & m  \\
V71      & RRab & 17.239 & 16.049 & 1.133 & 0.910 & 0.526707 & 8341.7790 & 17:37:28.62 & -03:15:35.6 & m  \\
V73      & SR/L?& \textit{14.669} & \textit{12.126} & -  & - &  --  & -  & 17:37:36.56 & -03:14:38.1 & m  \\
V74      & SR   & 14.433 & 11.810 & 0.223 & 0.115 & 20.39430 & 9078.1457 & 17:37:36.70 & -03:13:14.8 & m  \\
V75      & RRab & 16.686 & 15.264 & 0.809 & 0.460 & 0.545284 & 8659.7781 & 17:37:38.61 & -03:14:55.5 & m? \\
V76      & CW   & 15.93  & 14.710 & 0.720 & 0.471 & 1.8899   & 9078.2493 & 17:37:29.25 & -03:14:45.0 & m  \\
V77      & RRab & 17.115 & 15.822 & 0.481 & 0.385 & 0.792157 & 8630.7462 & 17:37:28.99 & -03:13:47.5 & m  \\
V78      & RRc  & 17.237 & 16.160 & 0.405 & 0.249 & 0.310265 & 8278.7969 & 17:37:27.16 & -03:14:48.5 & m  \\
V79      & RRab & 17.527 & 16.344 & 1.218 & 0.933 & 0.559898 & 9076.1118 & 17:37:35.53 & -03:15:00.4 & m  \\
V80      & RRc  & 17.091 & 16.081 & 0.426 & 0.257 & 0.315810 & 8341.7631 & 17:37:33.97 & -03:17:14.5 & m  \\
\hline
\end{tabular}
\center{\quad 
1. These values are intensity-weighted means, except when in italic font which are magnitude-weighted means.
2. All light curves in this paper are phased with these periods.
3. Membership status: m=member, f=field star, ?= no proper motion available, m/f and m/m: these stars are blends with individual likely membership status 4. Out of the field of our images. 5. Period from CP18. 6. Newly reported in the present work.}
\end{center}
\end{table*}

\begin{table*}
\small
\addtocounter{table}{-1}
\scriptsize
\begin{center}
\caption{Continue}
\begin{tabular}{llccccccccc}
\hline
ID & Variable & $<V>^1$ & $<I>^1$ & $A_V$ & $A_I$ & $P^2$ (days) &  HJD$_{\rm max}$ &  RA  &  Dec. & mem$^3$\\
  &  Type  & (mag) & (mag) & (mag) & (mag) & this work &  (2450000+)  & (J2000.0) & (J2000.0)&\\
\hline
V88      & RRc  & 17.320 & 16.192 & 0.454 & 0.286 & 0.313145 & 8341.7809 & 17:37:31.04 & -03:14:33.1 & m  \\
V90      & RRc  & 17.189 & 15.997 & 0.436 & 0.264 & 0.351248 & 8661.6935 & 17:37:33.70 & -03:15:16.4 & m  \\
V91      & RRc  & 17.267 & 16.307 & 0.402 & 0.268 & 0.261279 & 8314.8363 & 17:37:29.92 & -03:15:21.2 & m  \\
V92      & RRab & 17.473 & 16.144 & 0.591 & 0.392 & 0.656203 & 8665.7506 & 17:37:33.82 & -03:14:41.2 & m  \\
V95      & RRc  & 17.147 & 16.041 & 0.432 & 0.283 & 0.359465 & 9110.1765 & 17:37:24.41 & -03:17:29.4 & m  \\
V96      & RRc  & 17.334 & 16.325 & 0.353 & 0.250 & 0.252479 & 9078.2347 & 17:37:24.91 & -03:14:47.6 & m  \\
V97      & EC   & 15.100 & 13.753 & 0.459 & 0.453 & 0.377437 & 9110.1429 & 17:37:25.27 & -03:18:37.1 & f  \\
V98      & RRc  & 17.243 & 16.241 & 0.393 & 0.262 & 0.257776 & 8286.9367 & 17:37:25.98 & -03:12:49.4 & m  \\
V99      & SR   & 14.497 & 12.299 & 0.091 & 0.083 & 16.31970 & 8314.7231 & 17:37:26.76 & -03:14:47.6 & m? \\
V102     & SR   & \textit{14.584} & \textit{12.321} & -  & -  & -- & -   & 17:37:31.59 & -03:16:01.0 & m  \\
V104     & RRc  & 16.660 & 15.797 & 0.294 & 0.190 & 0.262201 & 9078.2493 & 17:37:32.87 & -03:14:56.3 & ?  \\
V105     & RRc  & 17.450 & 16.368 & 0.574 & 0.346 & 0.279389 & 8629.9516 & 17:37:33.13 & -03:14:03.7 & m  \\
V106     & RRab & 17.328 & 16.083 & 0.935 & 0.689 & 0.547026 & 8286.7880 & 17:37:33.47 & -03:14:46.8 & m  \\
V107     & RRc  & 17.319 & 16.368 & 0.458 & 0.181 & 0.294971 & 8662.9057 & 17:37:33.75 & -03:14:48.5 & m  \\
V110     & RRc  & 17.275 & 16.248 & 0.448 & 0.296 & 0.301227 & 8629.9588 & 17:37:33.86 & -03:16:10.2 & m  \\
V114     & RRc  & \textit{17.299} & \textit{16.170} & -& -& 0.332567 & - & 17:37:34.09 & -03:14:52.8 & ?  \\
V116     & RRc  & 17.392 & 16.448 & 0.363 & 0.236 & 0.252287 & 8314.7532 & 17:37:34.23 & -03:15:37.7 & m  \\
V117     & RRc  & 17.434 & 16.357 & 0.643 & 0.385 & 0.339468 & 8659.7868 & 17:37:34.28 & -03:14:36.4 & m  \\
V118     & RRc  & 16.641 & 15.774 & 0.282 & 0.215 & 0.380509 & 8632.8819 & 17:37:34.45 & -03:16:12.9 & m  \\
V119     & RRc  & 16.833 & 15.620 & 0.511 & 0.164 & 0.330066 & 8629.9559 & 17:37:34.64 & -03:14:51.7 & m/f\\
V120     & SR/L?& \textit{14.278} & \textit{11.749} & -    & -  & - & -  & 17:37:34.74 & -03:13:30.8 & m  \\
V121     & RRc  & -      & -      & -     & -     & -        & -         & 17:37:35.21 & -03:15:19.8 & m  \\
V122     & RRab & 16.695 & 15.471 & 0.519 & 0.358 & 0.558875 & 8661.6906 & 17:37:35.47 & -03:14:39.4 & m  \\
V123     & RRc  & 17.220 & 16.259 & 0.332 & 0.216 & 0.284461 & 8631.9399 & 17:37:35.73 & -03:15:44.1 & m  \\
V124     & RRab & 17.300 & 16.010 & 0.827 & 0.533 & 0.570026 & 8286.7919 & 17:37:35.92 & -03:15:27.5 & m  \\
V126     & RRc  & 15.823 & 14.457 & 0.153 & 0.075 & 0.294236 & 8659.8127 & 17:37:36.01 & -03:14:53.7 & m  \\
V127     & RRc  & -      & -      & -     & -     &       -  & -         & 17:37:36.09 & -03:14:33.0 & m  \\
V128     & RRc  & -      & -      & -     & -     & -        & -         & 17:37:36.11 & -03:13:52.9 & m  \\
V129     & RRc  & 17.099 & 16.045 & 0.273 & 0.166 & 0.279880 & 8314.7268 & 17:37:36.29 & -03:15:01.3 & m  \\
V130     & RRab & 17.262 & 15.956 & 0.744 & 0.499 & 0.595842 & 9078.1323 & 17:37:36.45 & -03:15:25.2 & m  \\
V131     & RRc  & 16.772 & 15.342 & 0.310 & 0.141 & 0.269686 & 9076.2727 & 17:37:36.45 & -03:14:54.3 & ?  \\
V132     & RRab & 17.360 & 16.085 & 1.127 & 0.749 & 0.482211 & 8659.7580 & 17:37:36.72 & -03:15:14.5 & m  \\
V133     & RRc  & 17.161 & 16.145 & 0.482 & 0.376 & 0.303017 & 8342.8060 & 17:37:36.87 & -03:18:15.1 & m  \\
V135     & RRc  & 17.366 & 16.185 & 0.441 & 0.327 & 0.338506 & 8659.8098 & 17:37:36.99 & -03:13:42.1 & m  \\
V136     & RRc  & 17.183 & 16.093 & 0.493 & 0.253 & 0.330080 & 8314.8740 & 17:37:37.02 & -03:15:24.1 & m  \\
V137     & RRc  & 17.061 & 16.101 & 0.395 & 0.244 & 0.276214 & 8278.8512 & 17:37:37.66 & -03:14:40.5 & ?  \\
V138     & RRc  & 16.949 & 15.869 & 0.408 & 0.250 & 0.371120 & 8631.9009 & 17:37:38.12 & -03:15:32.6 & m  \\
V139     & RRc  & 17.290 & 16.288 & 0.478 & 0.317 & 0.271854 & 9110.1429 & 17:37:38.21 & -03:17:22.9 & m  \\
V140     & RRab & 17.067 & 15.768 & 0.392 & 0.286 & 0.760854 & 8314.8155 & 17:37:38.40 & -03:15:51.4 & m  \\
V141     & RRab & 17.166 & 15.840 & 0.758 & 0.624 & 0.643125 & 9078.1510 & 17:37:38.48 & -03:14:27.0 & m  \\
V142     & RRab & 17.3   & 16.1   & 0.837 & 0.681 & 0.461696 & 9076.1978 & 17:37:38.89 & -03:14:01.9 & m  \\
V143     & RRc  & 17.291 & 16.204 & 0.406 & 0.286 & 0.271854 & 9110.1429 & 17:37:39.02 & -03:16:30.5 & m  \\
V144     & RRc  & -      & -      &   -   & -     & -        & -         & 17:37:39.44 & -03:14:29.3 & m  \\
V145     & RRc  & 17.080 & 16.052 & 0.380 & 0.244 & 0.300320 & 9134.0870 & 17:37:40.15 & -03:15:00.9 & m  \\
V147     & RRab & 17.147 & 15.728 & 1.063 & 0.638 & 0.492762 & 8278.7846 & 17:37:40.44 & -03:15:54.9 & m  \\
V148     & RRc  & 17.266 & 16.262 & 0.344 & 0.285 & 0.265167 & 8631.8576 & 17:37:40.93 & -03:16:59.2 & m  \\
V150     & RRab & 17.126 & 15.790 & 0.459 & 0.385 & 0.802765 & 8278.9405 & 17:37:41.31 & -03:14:07.5 & m  \\
V151     & RRc  & 17.432 & 16.293 & 0.472 & 0.307 & 0.319666 & 8344.7391 & 17:37:41.35 & -03:14:22.1 & m  \\
V152     & SR   & 14.338 & 11.853 & 0.196 & 0.147 & 38.54470 & 9110.1133 & 17:37:42.09 & -03:15:36.3 & m  \\
V153     & RRc  & 17.238 & 16.331 & 0.359 & 0.191 & 0.264831 & 8314.7759 & 17:37:42.12 & -03:11:55.8 & m  \\
V154     & RRc  & 17.399 & 16.375 & 0.448 & 0.312 & 0.254995 & 9110.1032 & 17:37:42.70 & -03:13:58.8 & m  \\
V155     & SR   & 14.663 & 12.544 & 0.080 & 0.071 & 36.54330 & 8662.9475 & 17:37:43.09 & -03:14:04.5 & m  \\
V157     & RRc  & 17.438 & 16.406 & 0.485 & 0.344 & 0.262598 & 8286.9089 & 17:37:43.97 & -03:16:12.2 & m  \\
V158     & RRab & 17.337 & 15.962 & 0.213 & 0.172 & 0.722078 & 8659.7580 & 17:37:44.01 & -03:13:00.2 & m  \\
V159     & RRc  & 17.344 & 16.328 & 0.458 & 0.291 & 0.262598 & 8286.9089 & 17:37:44.15 & -03:13:44.0 & m  \\
V160     & RRc  & 17.090 & 16.051 & 0.511 & 0.323 & 0.340272 & 8278.9069 & 17:37:44.42 & -03:15:33.6 & m  \\
V162     & SR   & 14.565 & 12.448 & 0.136 & 0.072 & 127.3005 & 8344.7896 & 17:37:49.81 & -03:11:48.1 & m  \\
V163$^4$ & RRc  & -      & -      & -     & -     & -        & -         & 17:37:53.87 & -03:14:16.6 & m  \\
V164$^4$ & RRc  & -      & -      & -     & -     & -        & -         & 17:37:56.25 & -03:10:15.4 & m  \\
V165     & SR   & \textit{14.918} & \textit{12.357} & - & -   & -   & -  & 17:37:36.94 & -03:14:53.2 & m  \\
V166     & RRab & 17.091 & 15.591 & 0.644 & 0.568 & 0.673005 & 8314.6928 & 17:37:36.61 & -03:14:17.9 & m  \\
V167     & CW   & 15.517 & 13.981 & 0.179 & 0.179 & 6.2010   & 8630.7462 & 17:37:33.47 & -03:15:27.9 & m  \\
V168     & EW   & \textit{16.564} & \textit{15.424} & -& -& 1.268656 & 4277.7523 & 17:37:26.12 & -03:09:45.5 & f  \\
V169     & RRc  & \textit{16.799} & \textit{15.672} & -& -& 0.3489$^5$   & - & 17:37:35.35 & -03:15:22.5 & m/m\\
V170     & SR   & 14.312 & 12.016 & 0.131 & 0.080 & 22.77620 & 8341.7556 & 17:37:41.50 & -03:16:03.3 & m  \\
V171     & SR   & 14.297 & 11.950 & 0.134 & 0.076 & 52.45570 & 9078.1583 & 17:37:39.84 & -03:15:06.7 & m  \\
V172     & RRab & 17.715 & 16.314 & 0.905 & 0.658 & 0.720411 & 8278.9319 & 17:37:33.95 & -03:15:11.8 & m  \\
\hline
\end{tabular}
\end{center}
\end{table*}

\begin{table*}
\small
\addtocounter{table}{-1}
\scriptsize
\begin{center}
\caption{Continue}
\begin{tabular}{llccccccccc}
\hline
ID & Variable & $<V>^1$ & $<I>^1$ & $A_V$ & $A_I$ & $P^2$ (days) &  HJD$_{\rm max}$ &  RA  &  Dec. & mem$^3$\\
  &  Type  & (mag) & (mag) & (mag) & (mag) & this work &  (2450000+)  & (J2000.0) & (J2000.0)&\\
\hline
V173$^6$ & RRc  & 17.252 & 16.272 & 0.209 & 0.139 & 0.253163 & 8314.6948 & 17:37:42.14 & -03:15:29.2 & m  \\
V174$^6$ & RRc  & 17.319 & 16.254 & 0.193 & 0.149 & 0.262191 & 8314.7212 & 17:37:23.24 & -03:14:51.9 & m  \\
         & -    &        &        &       &       & 0.256387 &           &             &             &    \\
V175$^6$ & RRc  & 17.491 & 16.260 & 0.660 & 0.416 & 0.299130 & 8344.7645 & 17:37:40.82 & -03:14:22.1 & m  \\
V176$^6$ & SR   & \textit{14.688} & \textit{12.476} & 0.100 & 0.082 & 11.21 & 8342.7890 & 17:37:47.85 & -03:16:36.6 & m  \\
V177$^6$ & SX   & 18.841 & 17.841 & 0.260 & 0.378 & 0.068984 & 8286.7880 & 17:37:37.91 & -03:16:47.9 & m  \\
         & -    &        &        &       &       & 0.053607 &           &             &             &    \\
V178$^6$ & SR   & \textit{15.246} & \textit{13.460} & 0.118 & 0.123 & 18.05 &    -      & 17:37:41.88 & -03:11:56.9 & m  \\
V179$^6$ & SR   & \textit{14.757} & \textit{12.581} & -     & -     &   -   & -         & 17:37:30.49 & -03:15:53.1 & m  \\
V180$^6$ & SR   & \textit{14.802} & \textit{12.583} & 0.081 & 0.048 & 15.76 & -         & 17:37:33.25 & -03:15:38.4 & m  \\
V181$^6$ & SR   & \textit{14.676} & \textit{12.490} & 0.064 & 0.047 & 18.42 & -         & 17:37:40.94 & -03:15:32.9 & m  \\
V182$^6$ & SR   & \textit{14.910} & \textit{12.923} & 0.126 & 0.107 & 85.26 & -         & 17:37:36.98 & -03:15:31.9 & m  \\
V183$^6$ & SR   & \textit{14.237} & \textit{11.968} & -     & -     &    -  & -         & 17:37:37.42 & -03:15:24.2 & m  \\
V184$^6$ & SR   & \textit{14.142} & \textit{12.098} & 0.104 & 0.102 & 123.07& -         & 17:37:38.77 & -03:15:18.3 & m  \\
V185$^6$ & SR   & \textit{14.359} & \textit{12.181} & 0.093 & 0.086 & 19.09 & -         & 17:37:35.98 & -03:15:15.0 & m  \\
V186$^6$ & SR   & \textit{14.821} & \textit{12.678} & -     & -     & 57.92 & -         & 17:37:34.97 & -03:15:04.1 & m  \\
V187$^6$ & SR   & \textit{14.777} & \textit{12.617} & 0.105 & 0.101 & 17.61 & 8343.7562 & 17:37:33.51 & -03:14:39.7 & m  \\
V188$^6$ & SR   & \textit{14.823} & \textit{12.609} & 0.115 & 0.081 & 18.07 & 9076.1342 & 17:37:38.38 & -03:14:37.4 & m  \\
V189$^6$ & SR   & \textit{14.483} & \textit{12.237} & 0.060 & -     & 22.54 & 8341.7902 & 17:37:28.99 & -03:14:37.0 & m  \\
V190$^6$ & SR   & \textit{14.742} & \textit{12.596} & -     & -     & 64.05 & -         & 17:37:34.98 & -03:14:34.9 & m  \\
V191$^6$ & SR   & \textit{14.339} & \textit{11.928} & 0.128 & 0.119 & 17.23 & 8342.7909 & 17:37:43.79 & -03:14:23.4 & m  \\
V192$^6$ & SR   & \textit{14.470} & \textit{12.106} & 0.171 & 0.124 & 16.69 & 8314.8476 & 17:37:32.65 & -03:14:12.9 & m  \\
V193$^6$ & SR   & \textit{15.065} & \textit{12.909} & 0.080 & 0.081 & 33.32 & 9078.1453 & 17:37:37.96 & -03:14:12.1 & m  \\
V194$^6$ & SR   & \textit{14.765} & \textit{12.549} & 0.122 & 0.080 & 32.46 & 8341.7940 & 17:37:38.64 & -03:13:44.7 & m  \\
\hline
\end{tabular}
\end{center}
\end{table*}

\begin{figure*}
\begin{center}
\includegraphics[width=17cm]{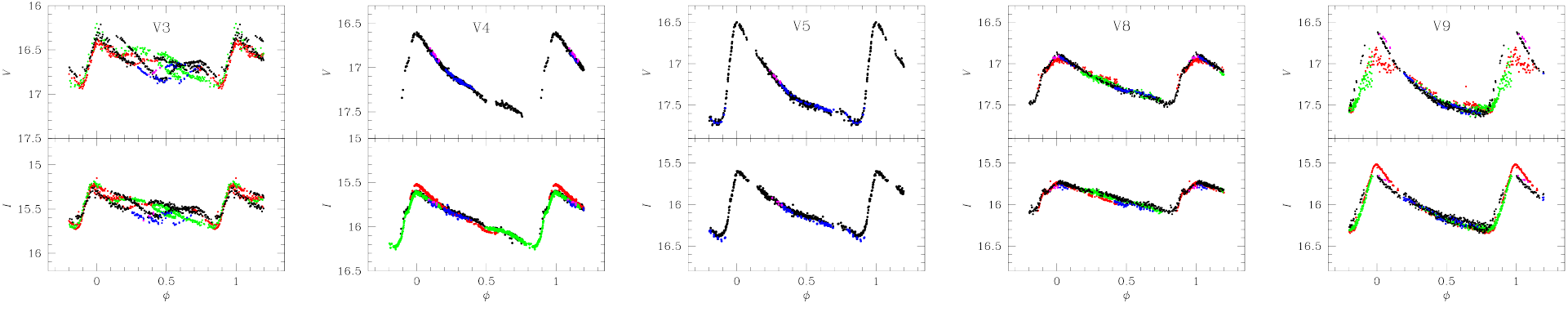}
\caption{Light curves of 54 RRab variables in M14. The symbols colour  colour code is; black for SPM 2018, red for SPM May 2019, green for SPM June 2019, blue for August-September IAO 2020 and lilac for October IAO 2020. The complete figure is available in electronic form.}
    \label{RRab1}
\end{center}
\end{figure*}

\begin{figure*}
\begin{center}
\includegraphics[width=17cm]{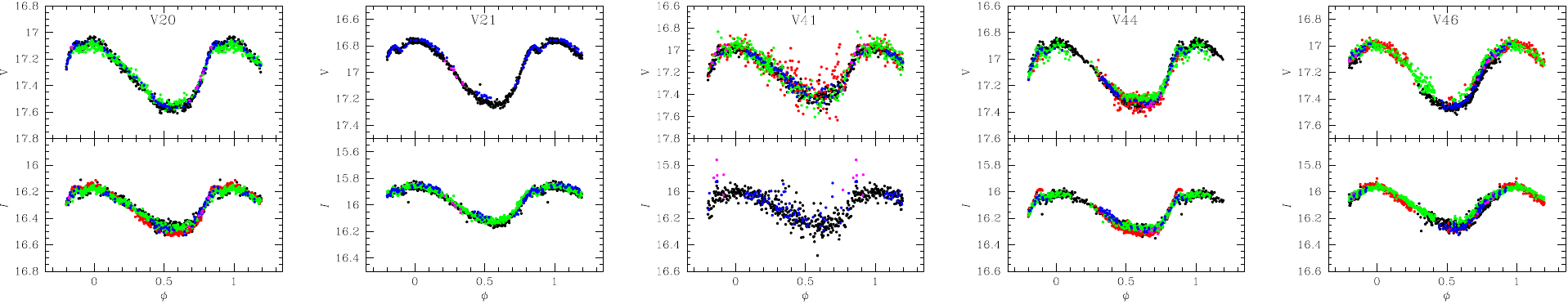}
\caption{Light curves of 45 RRc variables in M14. The colour code is the same as in Fig. \ref{RRab1}. The complete figure is available in electronic form.}
    \label{RRc1}
\end{center}
\end{figure*}

\begin{figure*}
\begin{center}
\includegraphics[width=15cm]{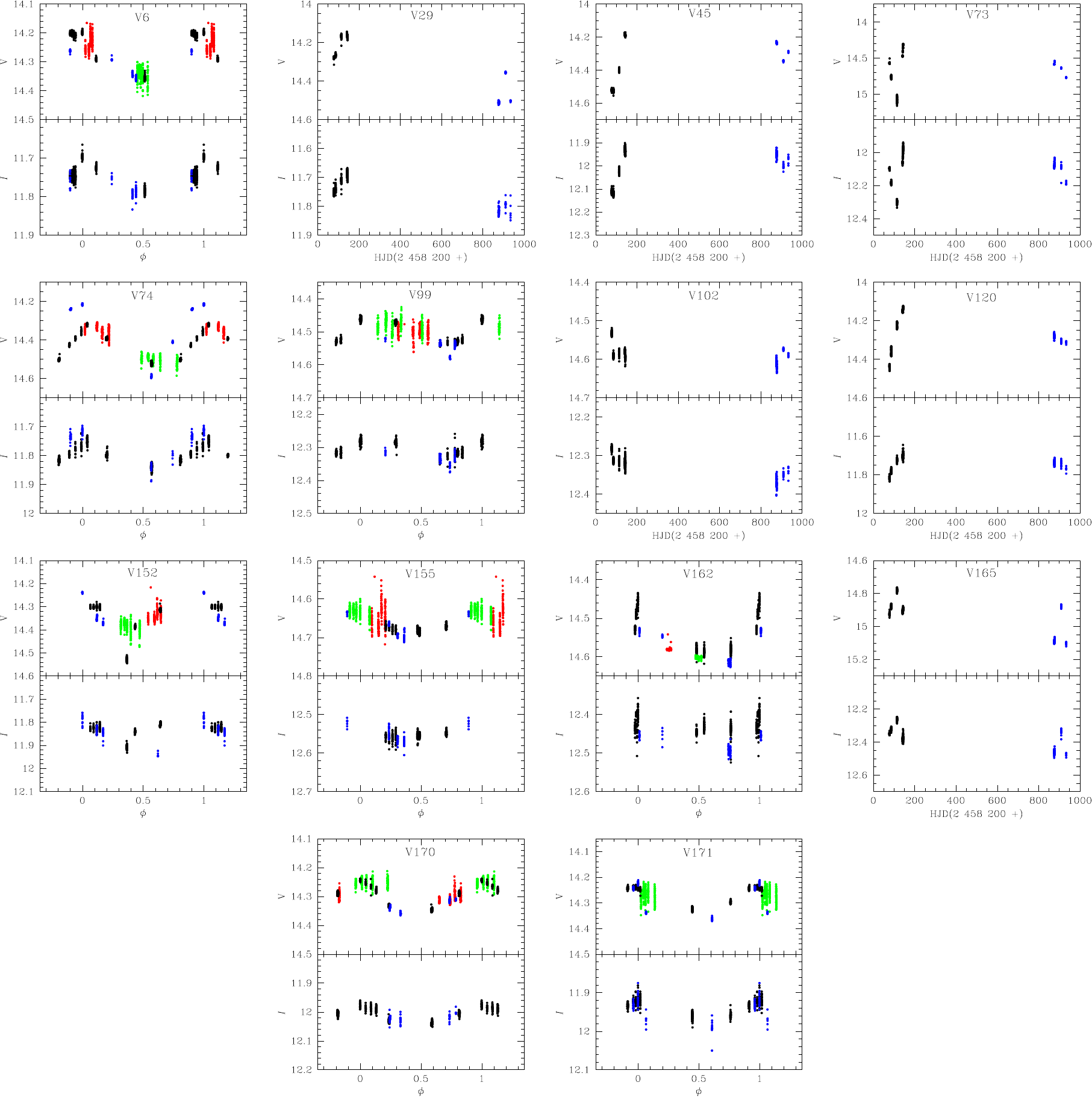}
\caption{Light curves of the long-period variables in M14. The colour code is the same as in Fig. \ref{RRab1}.}
    \label{SR_LC}
\end{center}
\end{figure*}

\begin{figure*}
\begin{center}
\includegraphics[width=14cm]{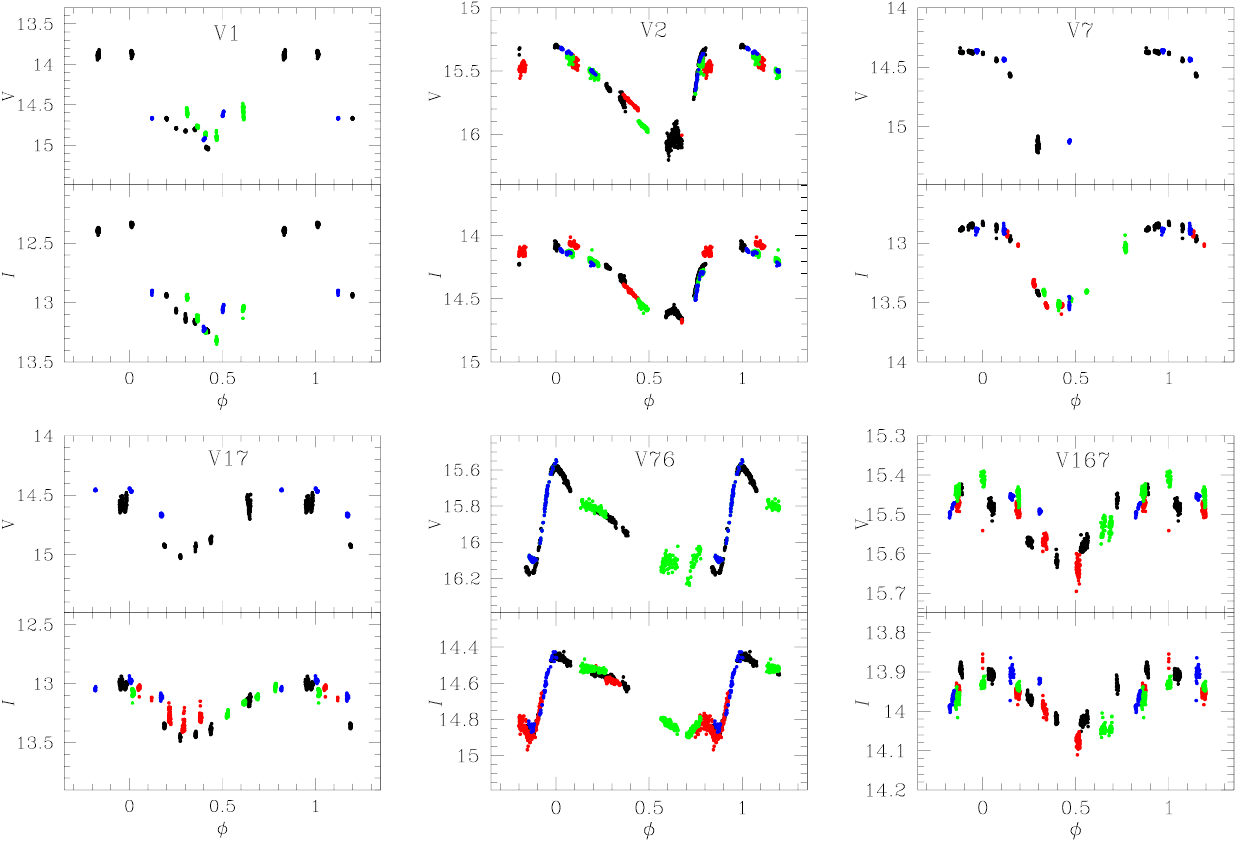}
\caption{Light curves of the CW variables in M14. The colour code is the same as in Fig. \ref{RRab1}.}
    \label{CW}
\end{center}
\end{figure*}

\begin{figure*}
\begin{center}
\includegraphics[width=13cm]{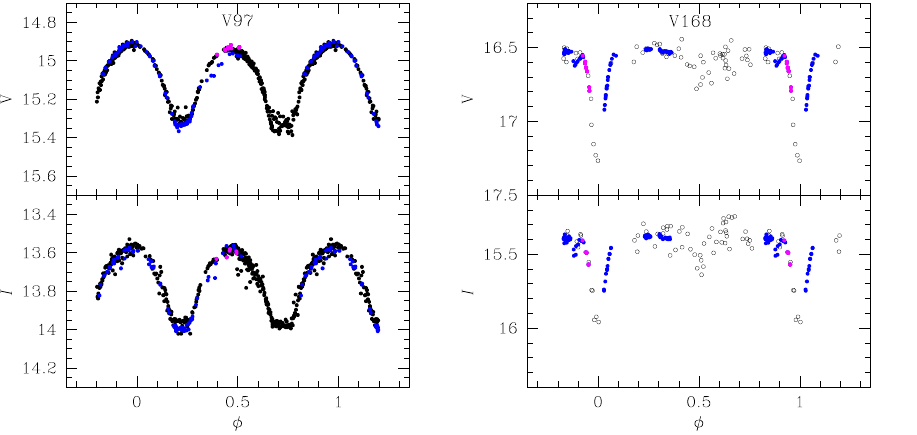}
\caption{Two eclipsing variables in M14. The colour code is the same as in Fig. \ref{RRab1}. For the case of V168 we included the data from CP18 (open circles), which allowed to complete the light curve.}
    \label{EC}
\end{center}
\end{figure*}

We were able to recover \emph{VI} light curves for 121 variables and their light curves are shown by types RRab, RRc, SR, CW and eclipsing binaries, in Figs. \ref{RRab1}, \ref{RRc1}, \ref{SR_LC}, \ref{CW} and \ref{EC}.

\subsection{Searching for new variables}

Though searching for new variables is not
the main aim of this work, we have taken advantage of the dense time-series of our images to perform a systematic search by a combination of different methods:\\

\begin{figure}
\begin{center}
\includegraphics[width=8cm]{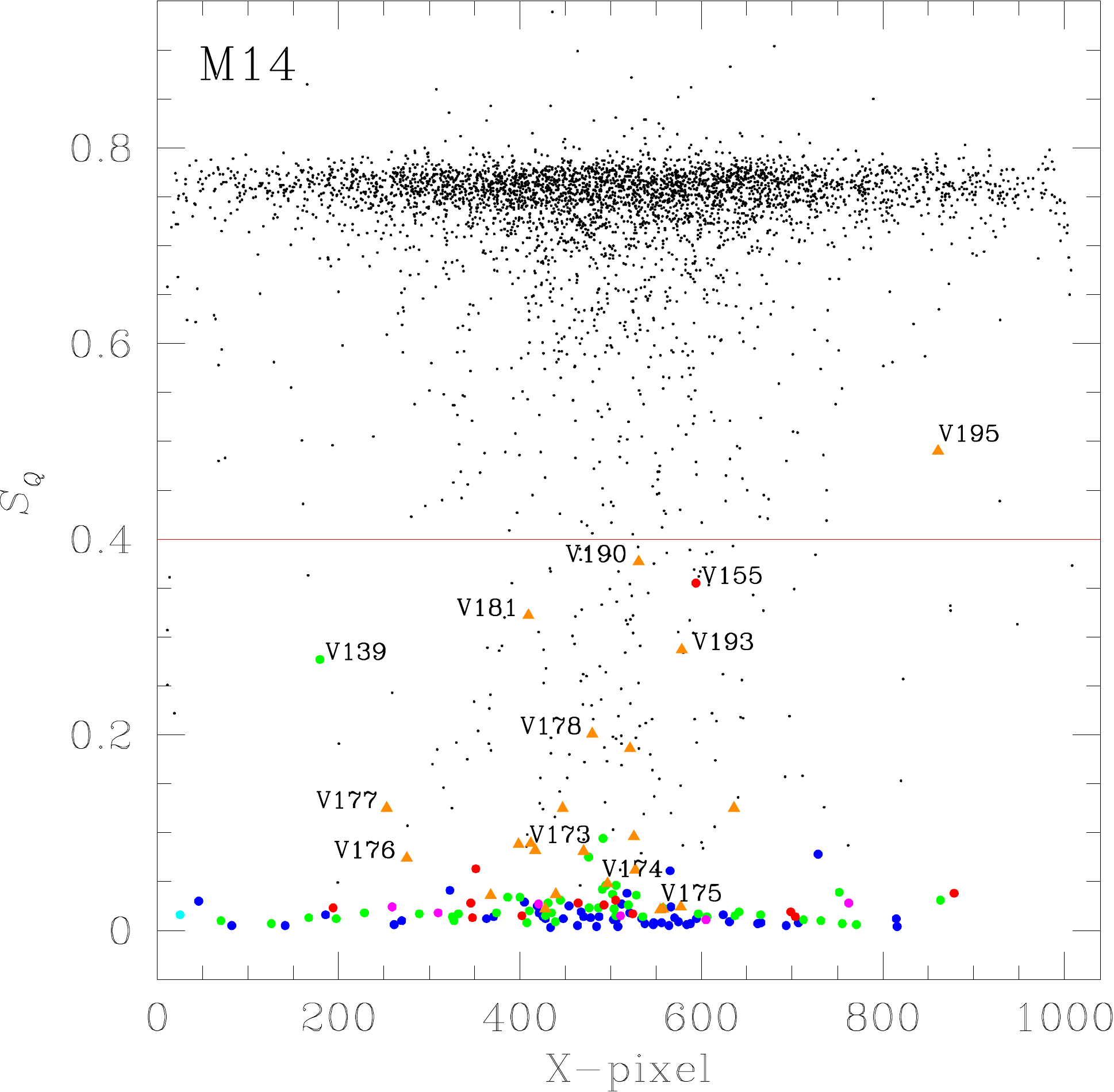}
\caption{The string-length $S_Q$ statistical parameter versus X-coordinate of the light source on the CCD, these data corresponds to SPM18. Blue, green, red, lilac and cyan points represent RRab, RRc, SR, CW and eclipsing binaries stars, respectively and orange triangles are newly discovered variable stars. The red line is an arbitrary threshold at $S_Q=0.4$. We visually inspected all the light curves with $S_Q$ below this value.}
    \label{SQ}
\end{center}
\end{figure}

\begin{enumerate}

\item String-length approach, in which an minimum statistical parameter $S_Q$, associated with the proper phasing of a light curve for a given a trial period is calculated. Stars with very small values of $S_Q$ are prone to be variables. Fig. \ref{SQ} shows the distribution of SQ's for all the stars measured in our light curve collection. We explored all the light curves with $S_Q$ below an arbitrary threshold of 0.4.
\item Isolation and subsequent exploration of the light curves contained in specific regions in the CMD, where it is usual to find variable stars, e.g. the HB, tip of the RGB and Blue Stragglers regions. 
\item Blinking of the residual images. This enables to distinguish and confirm the variability of new variables found by the other methods.
\end{enumerate}

The above three approaches have been described in detail by \citet{Yepez20}.

\begin{figure*}
\begin{center}
\includegraphics[width=17cm]{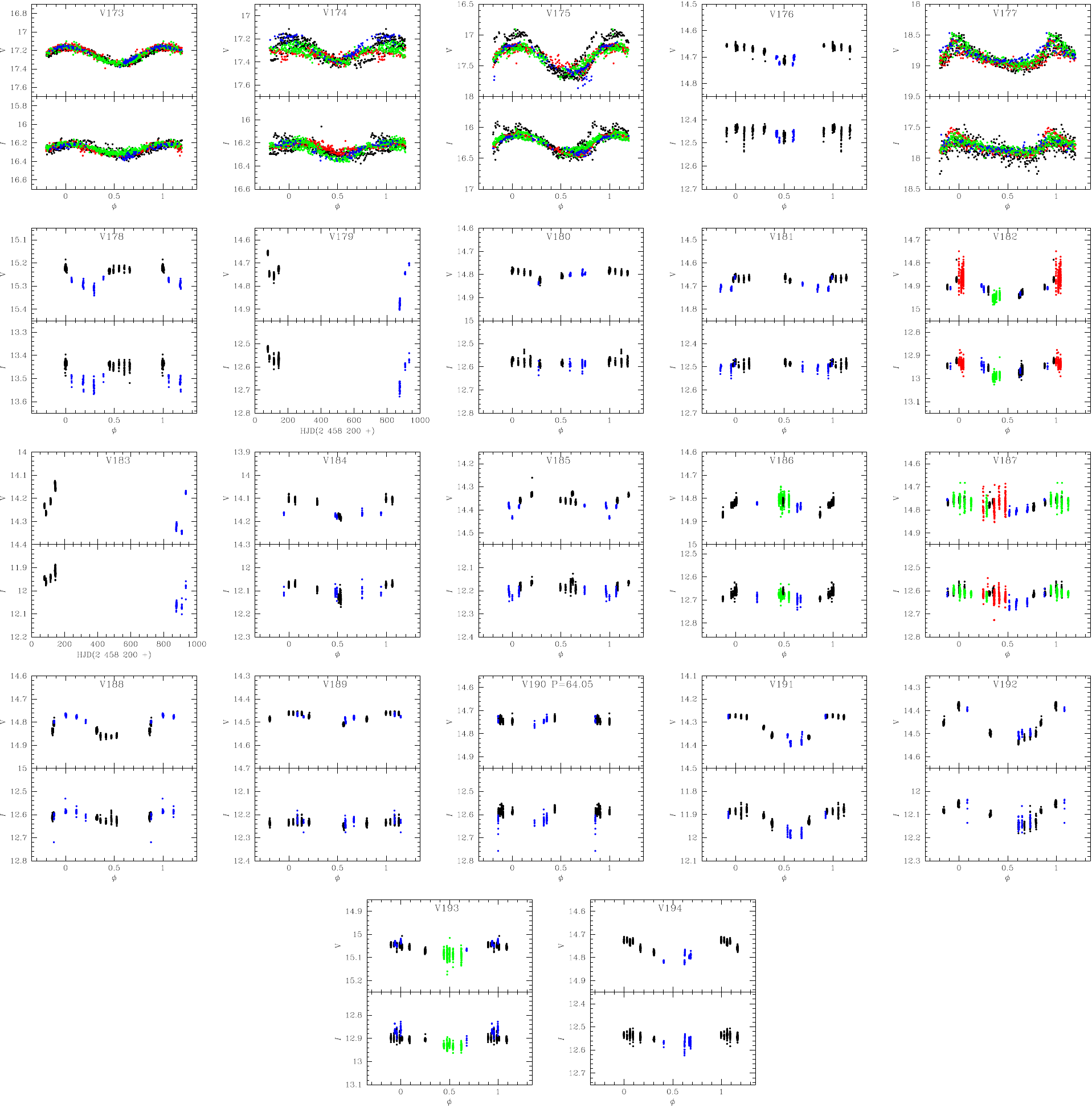}
\caption{Light curves of variables in M14 newly detected in this work. The colour code is the same as in Fig. \ref{RRab1}.}
    \label{News}
\end{center}
\end{figure*}

We found 3 new RRc stars, 1 of them with double mode ($\S$ \ref{doblemodo}), 18 SR or long-period variables and 1 SX Phe (the only one known in the cluster), their light curves are shown in the Fig. \ref{News}. In total 22 new variable stars are presently announced, for a new count of 155 authentic variables in M14: 55 RRab, 56 RRc, 1 RRd, 1 SX Phe, 6 type II Cepheids, 3 Eclipsing Binaries and 33 long-period or SR. The basic photometric and positional data of all these variables are listed in the Table \ref{tab:variables}. A discussion of the properties of peculiar or controversial stars is given in  Appendix \ref{AppA}.\\

\subsection{Double mode RR Lyrae stars}
\label{doblemodo}

\begin{figure*}
\begin{center}
\includegraphics[width=17.cm]{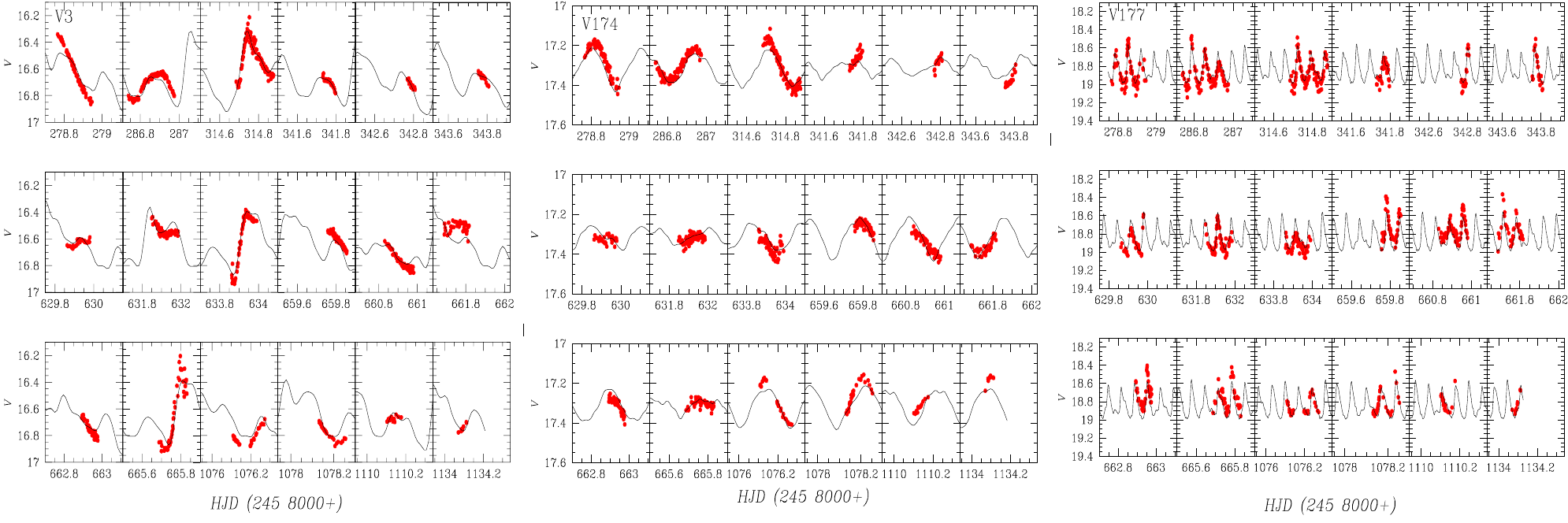}
\caption{The two-mode stars V3 (RRd), V174 (RRc with a non-radial mode) and V177 (SX Phe) fitted with a two period model. See $\S$ \ref{doblemodo} and \ref{SXPHE} for a discussion.}
    \label{Doublemode}
\end{center}
\end{figure*}

V3. This star is listed as RRab star in the CVSGC. However its light curve  suggests the presence of more than one period. The analysis using a simultaneous Fourier fit to the data reveals two periods: $P_0$=0.522380 and $P_1$=0.403379 for a ratio of $P_1/P_0$=0.77, which supports the idea that the star should be classified as a double mode or RRd. In Fig \ref{Doublemode} we show the data fitted by a double mode model of the above two frequencies. 

 V174 is a variable whose light curve shape and primary period suggest  a classification  as RRc, however, its light curve shows a large dispersion and amplitude modulations (Fig. \ref{News}) which can however be represented by double mode models as shown in Fig. \ref{Doublemode}. We explored the presence of a secondary period and found the periods 0.262191 d and 0.256387 d. The period ratio is 0.978, suggesting the presence of at least one non-radial mode.

\subsection{The SX Phe star V177}
\label{SXPHE}

No SX Phe stars were known in M14 previous to the present work. Our search has revealed the variable V177, whose period and position on the CMD support its classification as SX Phe. The light curve included in Fig. \ref{News}, displays substantial amplitude modulations, typical in multimode SX Phe. Two periods were identified: 0.068984 d and 0.053607 d, for a ratio of 0.78, which indicates that both modes are radial. The two-mode model fitting the data is shown in Fig. \ref{Doublemode}. 

\subsection{New SR variables in M14}
\label{SR}

The newly found SR variables were named V176 and V178-V194. In all cases, except for V179 and V183, we have been able to find a period. Their periods range between 11 and 123 days. Their light curves are shown in Fig. \ref{News}. 

\subsection{Eclipsing binaries in M14}

There are 3 known eclipsing binaries in M14, namely V28, V97 and V168, however V28  is out of the FoV of our images. The light curves of V97 and V168 are shown in Fig. \ref{EC}. V97 is clearly of the Algol type likely in contact, while V168 shows a deep eclipse and a mild suggestion of a secondary eclipse. We included for this star the data from CP18 after applying a small zero point shift, which helped to complete the curve and refine the period, given in Table \ref{tab:variables}.

\begin{figure*}
\begin{center}
\includegraphics[width=17cm,height=8.5cm]{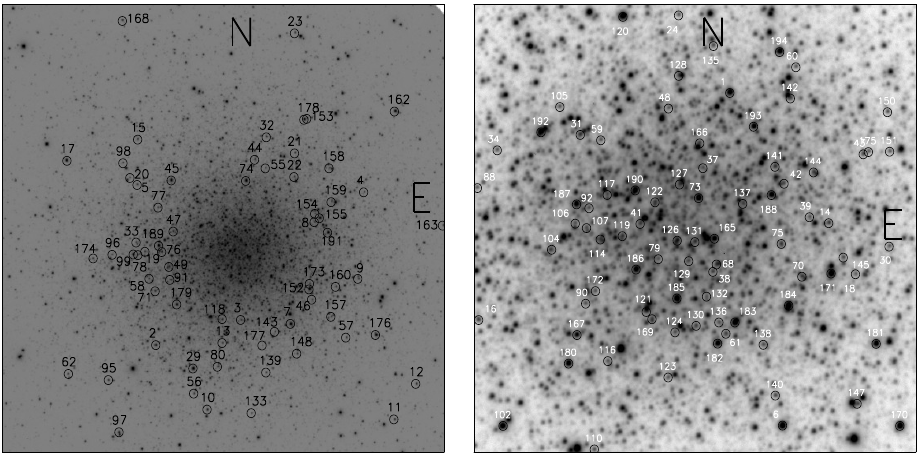}
\caption{Identification charts were built on the reference image for the Han20 data set. The field on the left is 9.7$\times$9.7 arcmin$^2$. The panel on the right contains the central region of the cluster and it is 2.75$\times$2.75 arcmin$^2$. Expansion of the digital version is recommended for clearness.}
    \label{IDcharts}
\end{center}
\end{figure*}

The identification charts with all variables included in this work are shown in Fig. \ref{IDcharts}. Their \emph{VI} photometric data are given in Table \ref{tab:vi_phot} of which we present only the small initial part. The complete table is available in electronic format at the Centre de Données astronomiques de Strasbourg (DCS) database. 

\begin{table}
\begin{center}
\caption{Time-series \textit{VI} photometry for the variables stars observed in this work$^*$}
\label{tab:vi_phot}
\centering
\begin{tabular}{cccccc}
\hline
Variable &Filter & HJD & $M_{\mbox{\scriptsize std}}$ &
$m_{\mbox{\scriptsize ins}}$
& $\sigma_{m}$ \\
Star ID  &    & (d) & (mag)     & (mag)   & (mag) \\
\hline
 V1 & $V$& 2458278.76587& 13.845 & 16.282 & 0.003 \\   
 V1 & $V$& 2458278.77078& 13.860  & 16.296 & 0.003 \\
\vdots   &  \vdots  & \vdots & \vdots & \vdots & \vdots  \\
 V1 & $I$ & 2458278.76669 & 12.328 & 15.049 & 0.002\\  
 V1 & $I$ & 2458278.76833 & 12.332 & 15.052 & 0.00  \\ 
\vdots   &  \vdots  & \vdots & \vdots & \vdots & \vdots  \\
 V2 & $V$ & 2458278.76587 & 15.723& 18.118 & 0.006 \\   
 V2 & $V$ & 2458278.77078 & 15.704& 18.100&  0.006 \\
\vdots   &  \vdots  & \vdots & \vdots & \vdots & \vdots  \\
 V2 & $I$ & 2458278.76669 & 14.477& 17.201 & 0.005 \\    
 V2 & $I$ & 2458278.76833 & 14.478&  17.202 & 0.005 \\   
\vdots   &  \vdots  & \vdots & \vdots & \vdots & \vdots  \\
\hline
\end{tabular}
\end{center}
* The standard and
instrumental magnitudes are listed in columns 4 and~5,
respectively, corresponding to the variable stars in column~1. Filter and epoch of
mid-exposure are listed in columns 2 and 3, respectively. The uncertainty on
$\mathrm{m}_\mathrm{ins}$, which also corresponds to the
uncertainty on $\mathrm{M}_\mathrm{std}$, is listed in column~6. A full version of this table is available at the CDS database.
\end{table}

\section{RR Lyrae Fourier decomposition and physical parameters}
\label{secFOURIER}
The light curves of RR Lyrae stars can be represented by a Fourier series of harmonics of the form:

\begin{equation}
    m(t) = A_{0} + \sum_{k=1}^{N}A_{k}cos(\frac{2\pi}{P}k(t-E_{0}) + \phi_{k}),
	\label{eq_foufit}
\end{equation}

\noindent
where $m(t)$ is the magnitude at time $t$, $P$ is the period in days and $E_0$ the epoch or a reference time of maximum light. A linear minimization routine is used to derive the amplitudes $A_k$ and phases $\phi_k$ of each harmonic, from which the Fourier parameters $\phi_{ij} = j\phi_{i} - i\phi_{j}$ and $R_{ij} = A_{i}/A_{j}$ are calculated, as well as the  intensity-weighted mean magnitudes $A_0 = <V>$.

Using the relevant Fourier parameters in the semi-empirical calibrations listed in Table \ref{tab:calib} for RRab and RRc stars, we can estimate individual physical parameters and mean values of [Fe/H] and distance for the parental cluster.

\begin{table*}
\begin{center}
\caption{Calibrations used to calculate [Fe/H] and $M_V$ for RR Lyrae stars from their light curve Fourier decomposition parameters.}
\label{tab:calib}
\begin{tabular}{lll}
\hline
Type & Calibration & Authors\\
\hline
     & ${\rm[Fe/H]_{Nem}}=-8.65-40.12P+5.96\phi_{31}^s+6.27\phi_{31}^sP-0.72(\phi_{31}^s)^2$ & \citet{Nemec2013}\\
RRab & ${\rm [Fe/H]_{\rm JK}}=-5.038-5.394P+1.345\phi_{31}^{(s)}$ & \citet{Jurcsik1996} \\
     & $M_V^{*}=-1.876\log P-1.158A_1+0.821A_3+0.41$ & \citet{Kovacs2001}\\
\hline
     & ${\rm [Fe/H]_{Nem}}=1.70-15.67P+0.20\phi_{31}^c-2.41\phi_{31}^cP+18.0P^2+0.17(\phi_{31}^c)^2$ & \citet{Nemec2013}\\
RRc  & ${\rm [Fe/H]}_{\rm ZW}=52.466P^2-30.075P+0.131\phi_{31}^{(c)2}+0.982\phi_{31}^{(c)}-4.198\phi_ { 31}^{(c)}P+2.424$ & \citet{Morgan2007} \\
     & $M_V^{*}=-0.961P-0.044\phi_{21}^{(s)}-4.447A_4+1.061$ & \citet{Kovacs1998} \\
\hline
\end{tabular}
\center{\quad * The zero point in these equations was calculated by \citet{Arellano2010}}
\end{center}
\end{table*}

For the RRab stars, we transformed [Fe/H]$_{\rm JK}$ in the scale of Jurcsik-Kovacs into [Fe/H]$_{\rm ZW}$ in the scale of \citet{Zinn1984}, by making use of  the relation [Fe/H]$_{\rm JK}$=1.431[Fe/H]$_{\rm ZW}$ + 0.88 \citep{Jurcsik1995}. In turn, the [Fe/H]$_{\rm ZW}$ values can be transformed into the UVES spectroscopic scale via the equation [Fe/H]$_{\rm UVES}$= $-0.413$ + 0.130~[Fe/H]$_{\rm ZW} - 0.356$~[Fe/H]$_{\rm ZW}^2$ \citep{Carretta2009}.

The values of  $T_{\rm eff}$, log$(L/{\rm L_{\odot}})$, $M/{\rm M_{\odot}}$ and $R/{\rm R_{\odot}}$ were calculated as in \citet{Arellano2010,Arellano2011}. We refer the interested reader to those papers.

The resulting physical parameters are reported in Table \ref{Tab:ParFis}. The values [Fe/H]$_{\rm Nem}$ and [Fe/H]$_{\rm UVES}$ are both in spectroscopic metallicity scale, and for the RRc stars the two values coincide as expected; however, in the case of the RRab stars a larger than expected difference is noted.

\begin{table*}
\scriptsize
\begin{center}
\caption{Physical parameters of RRab and RRc stars in M14 from their light curve Fourier decomposition.} 
\label{Tab:ParFis}

\begin{tabular}{cllllllllcc}
\hline
\multicolumn{9}{c}{RRab} \\
\hline 
ID&[Fe/H]$_{\rm ZW}$&[Fe/H]$_{\rm UVES}$ & [Fe/H]$_{\rm Nem}$ &$M_V$ & log~$T_{\rm eff}$  &log$(L/{L_{\odot}})$ &$M/{ M_{\odot}}$&$R/{ R_{\odot}}$ & $D(kpc)$&$D_m$\\
\hline
V4   & -1.28(6) & -1.16(6) & -0.75(10) & 0.461(10) & 3.811 & 1.720(4) & 0.63 & 5.80 &  9.55 & 2.5 \\
V5   & -1.44(2) & -1.34(2) & -1.21(4)  & 0.529(3)  & 3.814 & 1.697(1) & 0.73 & 5.57 &  9.97 & 1.2 \\
V8   & -1.30(4) & -1.18(4) & -0.66(5)  & 0.507(1)  & 3.801 & 1.702(1) & 0.63 & 5.96 &  9.57 & 2.2 \\
V10  & -1.39(2) & -1.28(3) & -1.09(4)  & 0.568(3)  & 3.810 & 1.680(1) & 0.67 & 5.57 &  8.74 & 2.1 \\
V12  & -1.13(7) & -1.01(6) & -0.65(10) & 0.613(9)  & 3.823 & 1.655(4) & 0.67 & 5.10 &  9.22 & 4.4 \\
V14  & -1.32(4) & -1.20(3) & -0.92(6)  & 0.589(6)  & 3.800 & 1.690(2) & 0.70 & 5.16 & 10.65 & 1.9 \\
V15  & -1.19(4) & -1.07(4) & -0.76(6)  & 0.566(4)  & 3.815 & 1.675(2) & 0.67 & 5.41 &  9.68 & 1.9 \\
V18  & -1.22(4) & -1.10(4) & -0.76(7)  & 0.576(7)  & 3.824 & 1.672(3) & 0.74 & 5.17 & 10.27 & 2.0 \\
V19  & -1.39(3) & -1.28(2) & -1.11(4)  & 0.495(3)  & 3.817 & 1.709(1) & 0.75 & 5.59 &  9.72 & 3.3 \\
V22  & -1.55(3) & -1.47(3) & -1.28(4)  & 0.473(1)  & 3.801 & 1.722(1) & 0.71 & 6.09 &  9.71 & 2.5 \\
V23  & -1.27(4) & -1.15(4) & -0.89(7)  & 0.525(4)  & 3.815 & 1.694(2) & 0.71 & 5.53 &  9.31 & 2.9 \\
V24  & -1.37(2) & -1.26(2) & -1.07(4)  & 0.560(3)  & 3.819 & 1.682(1) & 0.72 & 5.37 &  9.49 & 2.1 \\
V30  & -1.56(2) & -1.36(2) & -1.25(3)  & 0.547(3)  & 3.815 & 1.690(1) & 0.77 & 5.50 &  9.48 & 2.4 \\
V31  & -1.59(4) & -1.52(4) & -1.46(8)  & 0.520(4)  & 3.802 & 1.705(1) & 0.72 & 6.21 &  9.55 & 2.8 \\
V32  & -1.42(3) & -1.32(3) & -1.00(4)  & 0.483(1)  & 3.803 & 1.714(1) & 0.68 & 5.99 &  8.86 & 2.0 \\
V37  & -1.45(3) & -1.35(3) & -1.20(5)  & 0.707(3)  & 3.816 & 1.626(1) & 0.70 & 5.09 &  8.87 & 1.6 \\
V39  & -1.51(3) & -1.42(3) & -1.34(5)  & 0.465(4)  & 3.812 & 1.724(2) & 0.76 & 5.81 & 10.31 & 2.5 \\
V42  & -1.63(4) & -1.57(4) & -1.52(8)  & 0.559(3)  & 3.800 & 1.690(1) & 0.70 & 5.91 &  8.15 & 2.9 \\
V48  & -1.32(6) & -1.20(6) & -0.91(10) & 0.632(9)  & 3.822 & 1.652(3) & 0.75 & 5.11 & 10.97 & 2.6 \\
V92  & -1.28(7) & -1.17(6) & -0.74(11) & 0.517(4)  & 3.799 & 1.697(2) & 0.68 & 5.98 & 10.80 & 3.7 \\
V106 & -1.89(5) & -1.93(7) & -2.28(12) & 0.608(5)  & 3.806 & 1.679(2) & 0.78 & 5.68 &  9.69 & 2.7 \\
V132 & -1.42(7) & -1.31(7) & -1.12(14) & 0.636(13) & 3.820 & 1.653(5) & 0.73 & 5.15 &  9.92 & 3.5 \\
V141 & -1.66(4) & -1.61(5) & -1.57(8)  & 0.504(3)  & 3.802 & 1.713(1) & 0.71 & 6.01 &  9.34 & 1.8 \\
V147 & -1.37(8) & -1.26(8) & -1.05(14) & 0.667(12) & 3.818 & 1.640(5) & 0.69 & 5.11 &  8.86 & 4.0 \\
\hline
W. mean &-1.44(1) &-1.32(1)&-1.17(6) &0.527(1) &3.801(1) &1.689(1) &0.71 &5.56 & 9.61 \\
$\sigma$ &$\pm$0.17 &$\pm$0.19&$\pm$0.35 &$\pm$0.07 &$\pm$0.002 &$\pm0.027$ &$\pm$0.04 &$\pm$0.34 &$\pm$0.67 \\
\hline
\end{tabular}

\begin{tabular}{cllllllllc}
\hline
\multicolumn{9}{c}{RRc} \\
\hline 
ID&[Fe/H]$_{\rm ZW}$ & [Fe/H]$_{\rm UVES}$ & [Fe/H]$_{\rm Nem}$ &$M_V$ & log~$T_{\rm eff}$  &log$(L/{L_{\odot}})$ &$M/{ M_{\odot}}$&$R/{ R_{\odot}}$ & $D(kpc)$\\
\hline
V20  & -1.23(5)  & -1.11(5)  & -1.11(4)  & 0.601(5)  & 3.874 & 1.662(2) & 0.63 & 4.07 &  9.89 \\
V21  & -1.26(4)  & -1.15(4)  & -1.14(4)  & 0.538(5)  & 3.868 & 1.688(2) & 0.55 & 4.30 &  8.58 \\
V41  & -1.18(10) & -1.06(10) & -1.06(8)  & 0.585(14) & 3.875 & 1.667(5) & 0.65 & 4.07 &  9.29 \\
V44  & -1.27(4)  & -1.15(4)  & -1.16(4)  & 0.574(5)  & 3.871 & 1.674(2) & 0.59 & 4.18 &  9.00 \\
V46  & -1.50(12) & -1.41(15) & -1.40(15) & 0.546(9)  & 3.865 & 1.691(4) & 0.54 & 4.39 &  9.48 \\
V55  & -1.62(8)  & -1.56(10) & -1.55(9)  & 0.587(15) & 3.863 & 1.679(6) & 0.52 & 4.36 &  9.25 \\
V56  & -1.52(4)  & -1.43(5)  & -1.42(5)  & 0.519(5)  & 3.864 & 1.703(2) & 0.54 & 4.46 &  8.94 \\
V78  & -1.20(6)  & -1.08(6)  & -1.07(7)  & 0.582(6)  & 3.870 & 1.669(2) & 0.53 & 4.18 &  9.23 \\
V80  & -1.11(9)  & -1.00(8)  & -0.97(10) & 0.558(5)  & 3.870 & 1.677(2) & 0.53 & 4.22 &  9.03 \\
V88  & -1.32(6)  & -1.20(6)  & -1.20(6)  & 0.569(9)  & 3.868 & 1.677(4) & 0.54 & 4.25 &  9.66 \\
V90  & -1.18(11) & -1.06(11) & -0.98(14) & 0.527(12) & 3.866 & 1.691(5) & 0.49 & 4.36 &  9.43 \\
V91  & -1.15(12) & -1.03(11) & -1.04(10) & 0.653(5)  & 3.875 & 1.639(2) & 0.59 & 3.94 &  9.23 \\
V95  & -1.59(5)  & -1.52(6)  & -1.51(7)  & 0.502(5)  & 3.862 & 1.712(2) & 0.53 & 4.55 &  9.15 \\
V96  & -0.81(15) & -0.75(11) & -0.75(13) & 0.663(5)  & 3.879 & 1.629(2) & 0.58 & 3.83 &  9.50 \\
V98  & -0.90(13) & -0.82(10) & -0.82(12) & 0.654(9)  & 3.877 & 1.634(4) & 0.58 & 3.88 &  9.27 \\
V105 & -1.31(5)  & -1.20(6)  & -1.19(5)  & 0.563(9)  & 3.871 & 1.680(4) & 0.62 & 4.20 & 10.46 \\
V107 & -1.42(6)  & -1.31(7)  & -1.31(6)  & 0.564(9)  & 3.869 & 1.682(4) & 0.60 & 4.26 &  9.75 \\
V110 & -1.23(6)  & -1.11(6)  & -1.11(6)  & 0.609(5)  & 3.870 & 1.659(2) & 0.54 & 4.12 &  9.64 \\
V116 & -1.04(21) & -0.93(18) & -0.94(17) & 0.661(9)  & 3.877 & 1.634(4) & 0.60 & 3.88 &  9.64 \\
V117 & -1.42(10) & -1.32(11) & -1.31(12) & 0.532(19) & 3.865 & 1.695(7) & 0.53 & 4.40 & 10.36 \\
V123 & -1.33(14) & -1.21(16) & -1.21(14) & 0.586(10) & 3.871 & 1.671(4) & 0.60 & 4.17 &  9.22 \\
V129 & -0.91(13) & -0.82(10) & -0.81(13) & 0.635(6)  & 3.875 & 1.642(3) & 0.54 & 3.96 &  8.52 \\
V133 & -1.31(4)  & -1.20(5)  & -1.20(4)  & 0.565(5)  & 3.869 & 1.679(2) & 0.57 & 4.24 &  9.07 \\
V135 & -1.31(8)  & -1.19(8)  & -1.16(10) & 0.548(6)  & 3.866 & 1.685(2) & 0.51 & 4.33 &  9.96 \\
V136 & -1.10(7)  & -0.99(6)  & -0.93(8)  & 0.535(10) & 3.869 & 1.686(4) & 0.51 & 4.28 &  9.57 \\
V137 & -1.13(7)  & -1.01(7)  & -1.02(7)  & 0.607(9)  & 3.873 & 1.658(4) & 0.58 & 4.05 &  8.49 \\
V138 & -1.60(4)  & -1.53(5)  & -1.53(6)  & 0.515(6)  & 3.861 & 1.707(2) & 0.50 & 4.55 &  8.68 \\
V139 & -1.21(7)  & -1.09(7)  & -1.10(7)  & 0.614(9)  & 3.873 & 1.657(4) & 0.60 & 4.05 &  9.69 \\
V143 & -0.95(10) & -0.86(8)  & -0.78(12) & 0.598(6)  & 3.871 & 1.657(2) & 0.49 & 4.10 &  9.76 \\
V145 & -1.20(9)  & -1.08(9)  & -1.08(10) & 0.603(6)  & 3.870 & 1.661(2) & 0.54 & 4.13 &  8.85 \\
V148 & -1.00(17) & -0.90(14) & -0.91(15) & 0.661(5)  & 3.876 & 1.633(2) & 0.56 & 3.90 &  9.03 \\
V154 & -0.88(15) & -0.80(11) & -0.80(13) & 0.636(5)  & 3.878 & 1.641(2) & 0.59 & 3.90 &  9.78 \\
V157 & -1.11(5)  & -0.99(5)  & -1.00(5)  & 0.602(5)  & 3.875 & 1.659(2) & 0.62 & 4.03 & 10.05 \\
V159 & -1.20(7)  & -1.08(7)  & -1.09(7)  & 0.563(9)  & 3.872 & 1.676(4) & 0.57 & 4.18 &  9.86 \\
V160 & -1.49(4)  & -1.40(4)  & -1.39(5)  & 0.507(5)  & 3.864 & 1.707(2) & 0.55 & 4.47 &  9.29 \\
V173 &    --     &    --     &     --    & 0.637(14) & 3.879 & 1.638(5) & 0.58 & 3.86 &  9.43 \\
\hline
Weighted mean &-1.23(2) & -1.12(2) & -1.12(2) & 0.584(1) &3.858(1) &1.666(1) &0.61 &4.32 & 9.39 \\
$\sigma$ &$\pm$0.21 &$\pm$0.21 &$\pm$0.22 &$\pm$0.047 &$\pm$0.018 &$\pm$0.019 &$\pm$0.12 &$\pm$0.44 &$\pm$0.48 \\
\hline
\end{tabular}
\end{center}
\end{table*}

\section{The Oosterhoff type and the period-amplitude diagram of M14}

\begin{figure}
\begin{center}
\includegraphics[width=8cm]{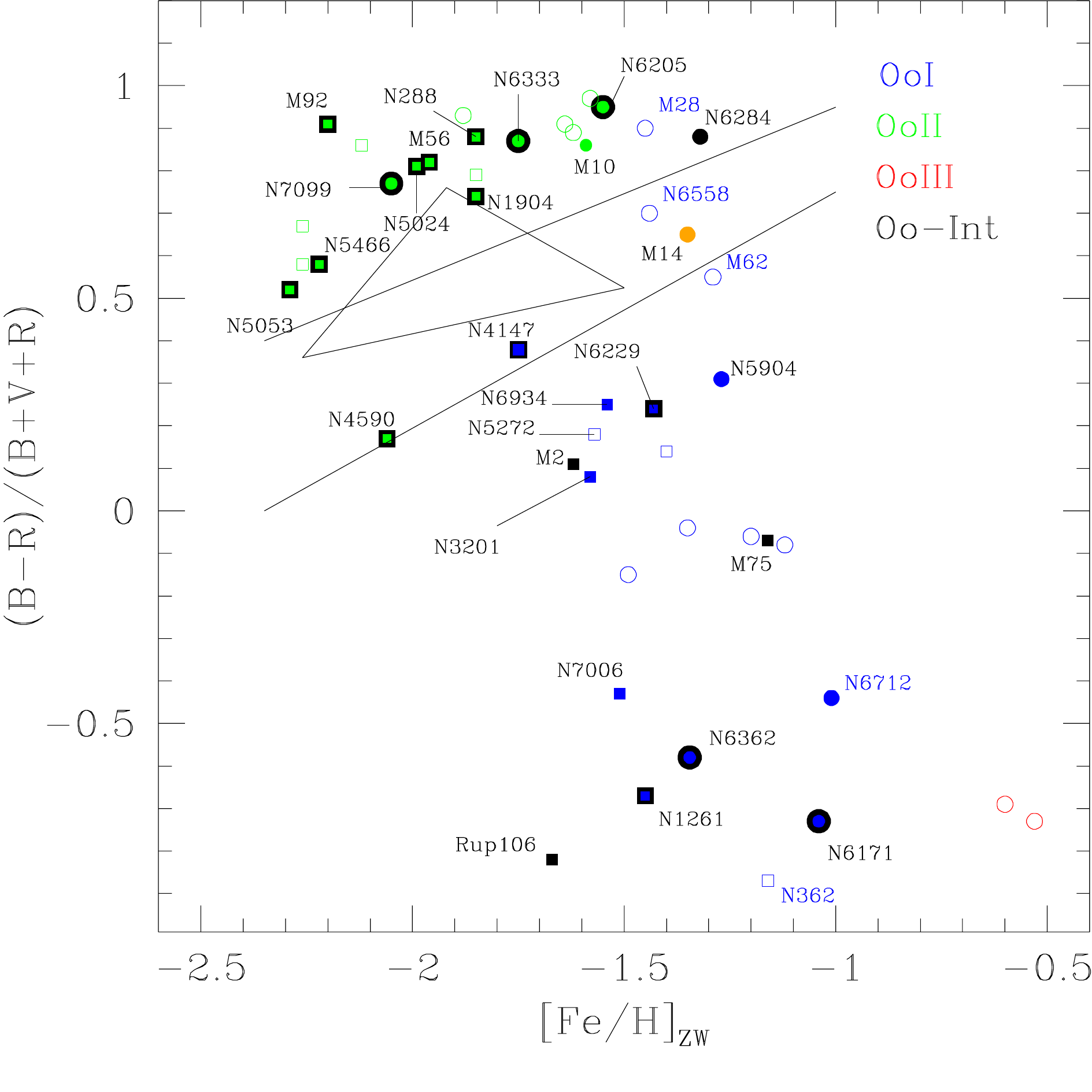}
\caption{Metallicity vs the HB structural parameter $\cal{L}$. Symbols with a black ribbon indicate clusters in which the RRc and RRab stars are well separated and do not share the bimodal region as opposed to the non-ribboned symbols. The triangle is the region identified by \citet{Catelan2009} as void of Galactic globular clusters  but preferentially populated by extragalactic globular clusters. The two straight lines delimit the Osterhoff gap according to the models of \citet{Bono1994}. Empty symbols are clusters not yet studied by our group.}
    \label{MetHB}
\end{center}
\end{figure}

It has been argued by \citet{Catelan2009}, that the RRab stars average period in Galactic globular clusters tend to avoid the range 0.58-0.62 days, the so called Oosterhoff gap, while in the case of extragalactic globulars, this period interval is rather preferentially occupied. \citet{Catelan2009} has also identified, on the HB structure parameter vs. metallicity plane, i.e. [Fe/H]-$\cal{L}$, a region void of Galactic but preferentially populated by extragalactic globular clusters, the triangle in Fig. \ref{MetHB}. The Oosterhoff type of M14 is of relevance for it has been suspected to be of extragalactic origin, \cite[e.g.][]{Gao2007}. Judging by the mean period of RRab stars, the relative number of RRc stars and the appearance of the period-amplitude diagram, or Bailey diagram, CP18 have argued in favour of an Oosterhoff intermediate classification, Oo-int. 

\begin{figure}
\begin{center}
\includegraphics[width=7cm]{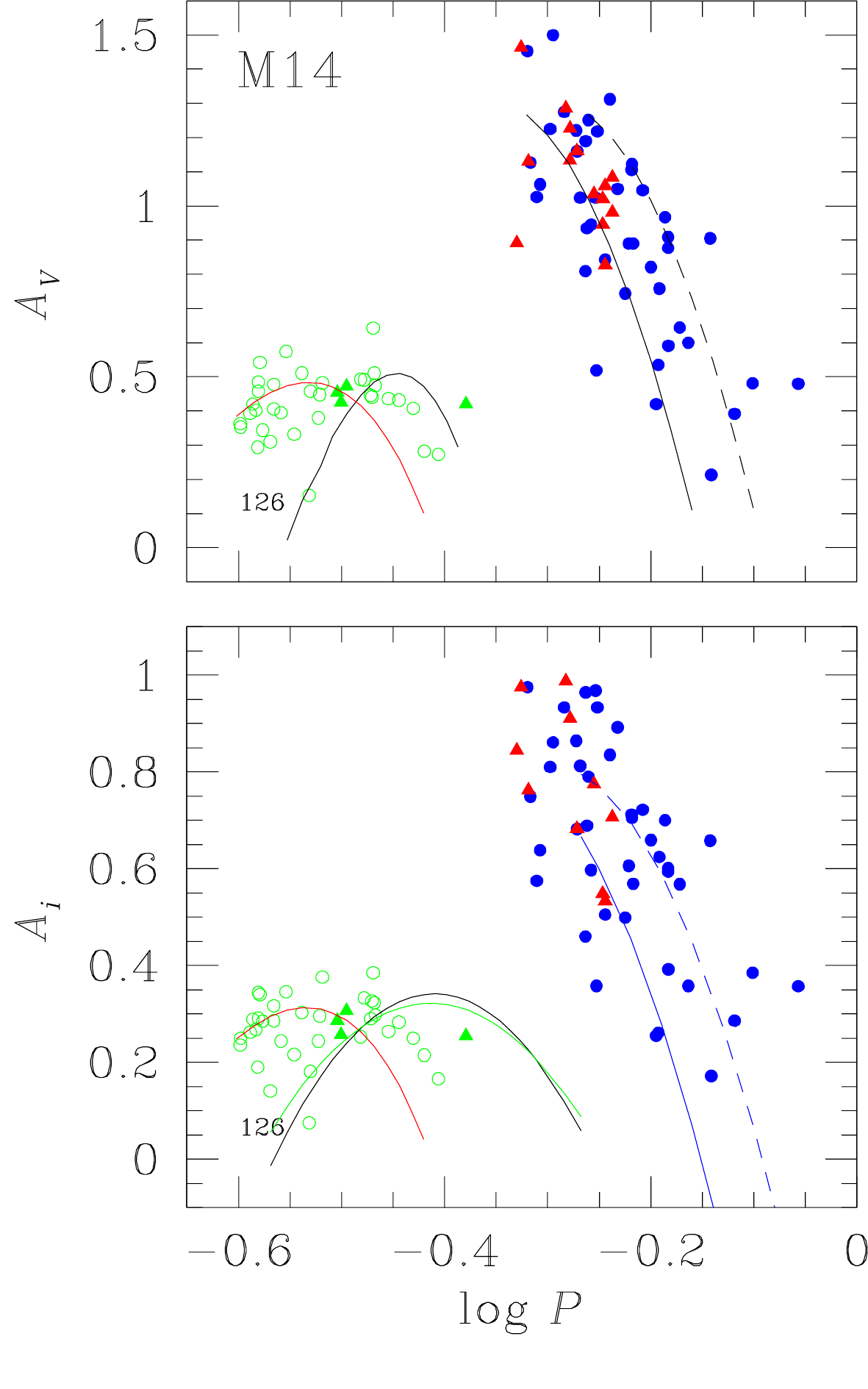}
\caption{The Period-Amplitude diagram of M14 in $V$ and $I$ band passes. Blue and green symbols corresponds to RRab and RRc stars, and red triangles represent stars with Blazhko-like modulations.}
    \label{PerAmp}
\end{center}
\end{figure}

M14 has commonly been referred to as an OoI cluster. From our data in Table \ref{tab:variables}, the average period of 43 RRab stars with no Blazhko modulations is  $<P_{ab}>=0.599 \pm 0.084$ d; the ratio of RRc stars relative to the total number of RR Lyrae $f_c$, including the new findings and ignoring those that seem to be field stars is $f_c=N_c/(N_c+N_{ab}) = 0.45$, a rather consistent value for an OoI cluster . Regarding the Bailey diagram, our version in Fig. \ref{PerAmp} is rather indistinguishable from that of CP18, and shows a large dispersion in the distribution of both RRab and RRc stars. Even ignoring the Blazhko amplitude modulated stars (red triangles in the figure) the stars neither conform to the typical OoI nor OoII reference loci. Thus, our results support the classification of M14 as Oo-int cluster.

\begin{figure*}
\begin{center}
\includegraphics[width=16.0cm]{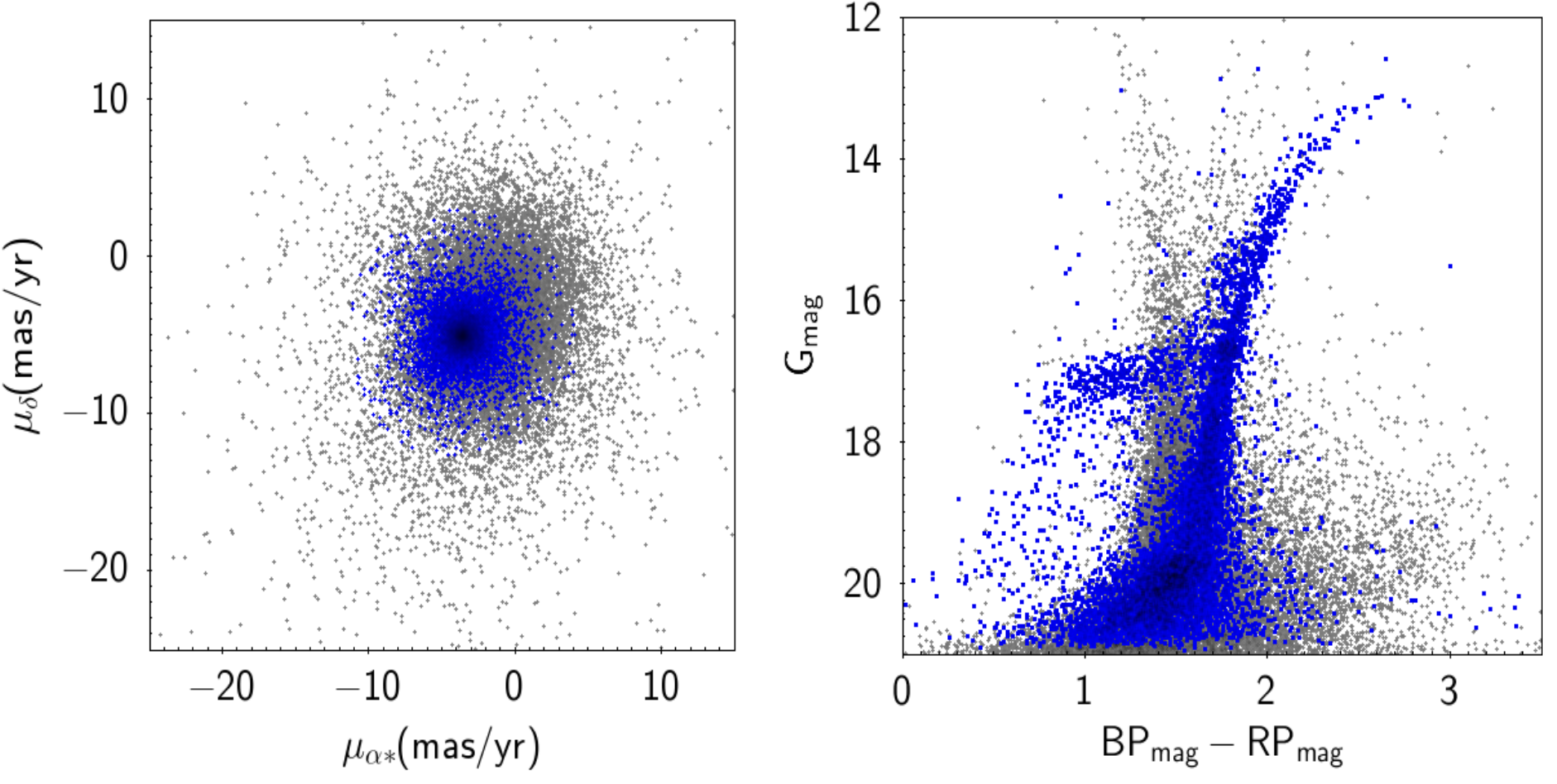}
\caption{The left panel shows the VPD of M14. Blue and gray dots represent cluster member and field stars respectively (see $\S$ \ref{gaia}). The right panel shows the corresponding CMD.}
    \label{VPD}
\end{center}
\end{figure*}

\begin{figure*}
\begin{center}
\includegraphics[width=17cm]{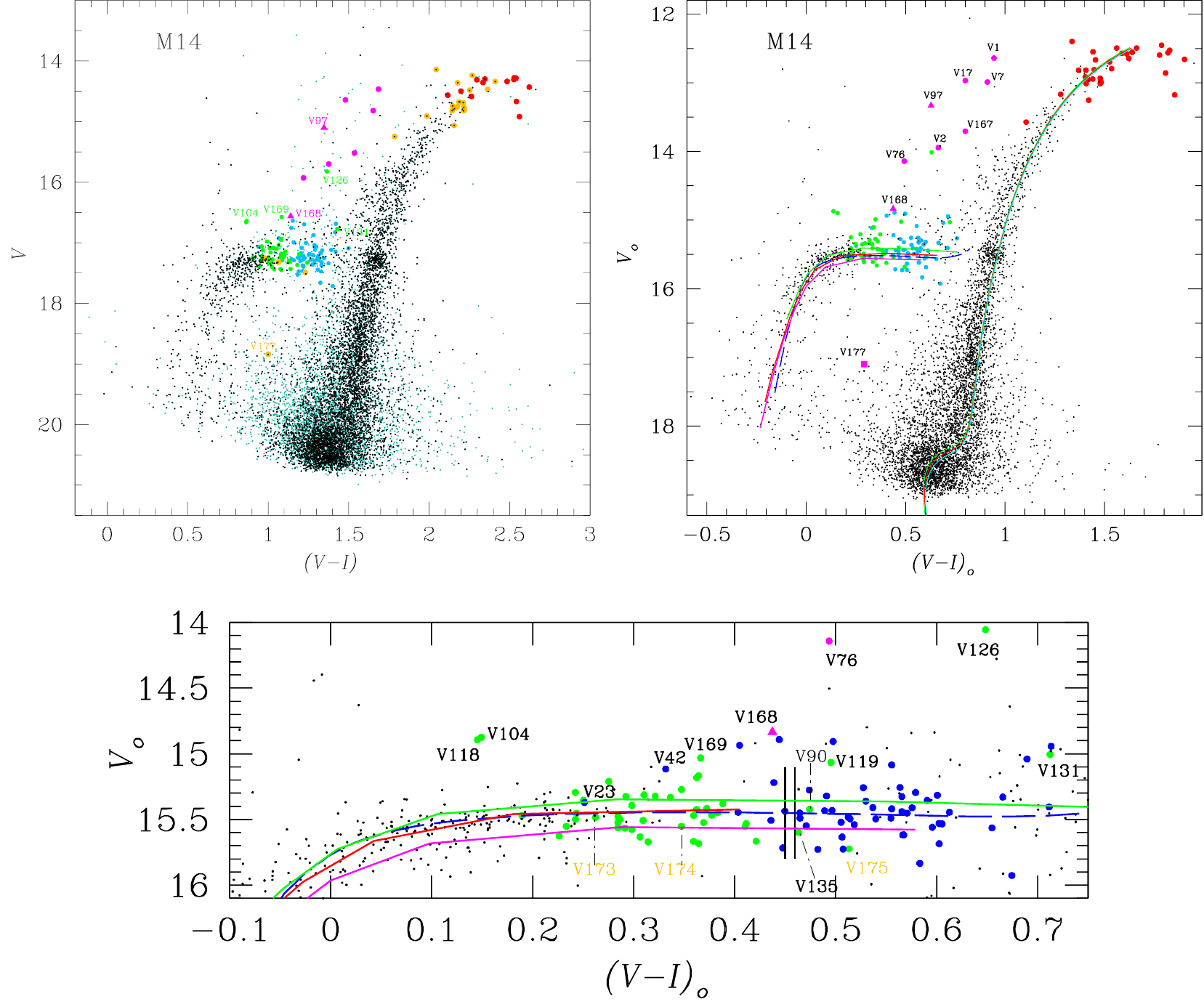}
\caption{M14 CMD and the HB region. The upper left panel shows the observed CMD including all the stars measured in the field of the cluster. Black and light blue small dots represent cluster member and field stars respectively, selected as explained in $\S$ \ref{gaia}. The variable stars colour code is: blue, green, red, lilac points, and lilac triangles for RRab, RRc, SR, CW, and eclipsing binary stars, respectively and orange points for newly discovered variables. The panel to the right is the CMD dereddened using the differential reddening map of CP13 as explained in $\S$ \ref{reddening}. The three isochrones of 12.0, 12.5 and 13.0 Gyrs, and [Fe/H] =$-1.3$,Y= 0.25, and [$\alpha$/Fe] = +0.4 and a ZAHB for [Fe/H] = $-1.4$ (segmented blue line), were calculated using the models from \citet{Vandenberg2014}. The continuous ZAHBs, built from the Eggleton code \citep{Pols1997, Pols1998, KPS1997} as described in $\S$ \ref{secCMD} are lilac, red and green for core masses on 0.48, 0.49 and 0.50$M_{\odot}$ respectively. All theoretical models have been placed at a distance of 9.3 kpc. The bottom panel is an expansion of the HB region where the empirical range for the first overtone red edge (FORE) is shown as two vertical black lines. (See $\S$ \ref{secCMD} for a discussion).}
    \label{CMD}
\end{center}
\end{figure*}

As for the HB structure goes, Fig. \ref{MetHB} shows the metallicity [Fe/H] versus the HB parameter $\cal{L} \equiv (B-R)/(B+V+R)$, where $\cal B,V$ and $\cal R$ refer to the number of stars to the blue, inside and to the red of the instability strip. The two straight lines in the figure delimit the Osterhoff gap according to the modelling of \citet{Bono1994}. An estimation of $\cal{L}$ from star member counts on our DCM (Fig. \ref{CMD}) and from that of $Gaia$ DCM (Fig. \ref{VPD}), yields a value 0.65 which along with the average [Fe/H]$_{\rm ZW}$  from the Fourier decomposition (Table \ref{Tab:ParFis}), places the cluster in the Oosterhoff gap in Fig. \ref{MetHB}. However, in spite of its Oo-int type, M14 falls outside the triangular zone generally preferred by the extragalactic clusters. In fact none of the other Oo-int clusters plotted as black symbols in Fig. \ref{MetHB} (NGC 6284, M2, M75 and Rup 106) is neither in the Oosterhoff gap nor the triangular zone. This has been noticed by \citet{Catelan2009} for NGC 6284, M75 and Rup 106 whom highlights some peculiarities in each case; scarce number of RR Lyrae, multimodal HB and lack of RRc stars respectively. While we may not claim similar peculiarities for M14, it is true that its CMD shows a large spread in both the RGB and HB. If this is due to the presence of more than one stellar generation and the consequent spread in He abundance \citep[e.g.][]{Milone2018}, the true value of  $\cal{L}$ would be unknown until the stars belonging to each generation are identified. However, these arguments may not be sufficient to consider M14 as of extragalactic origin. From the kinematical analysis of \citet{Massari2019}, M14 is not associated to any known merger event and is found to belong to a group of low-energy, highly bound, clusters probably formed {\it in-situ}.

We remark that the pulsating either-or region in this cluster, to the blue of the first overtone red edge (FORE) of the instability strip in the DCM of Fig. \ref{CMD}, is clearly shared by some RRc and RRab stars, as it happens in many OoI and apparently never in an OoII cluster.

\section{Distance to M14 from its variable stars}

\subsection{From the RR Lyrae P-L ({\it I}) relation}

Taking advantage of the numerous RR Lyrae in the cluster, an independent estimation of the cluster distance can be achieved via the P-L luminosity calibration in the $I$ band of \citet{Catelan2004}, which is of the form $M_I = 0.471-1.132~ {\rm log}~P +0.205~ {\rm log}~Z$, with ${\rm log}~Z = [M/H]-1.765$; $[M/H] = \rm{[Fe/H]} - \rm {log} (0.638~f + 0.362)$ and log~f = [$\alpha$/Fe]. We adopted [$\alpha$/Fe]=+0.3 \citep{Sal93}. In the above expression $P$ should be that of the fundamental mode, hence for the RRc stars, their periods were 'fundamentalised' by assuming the ratio $P1/P0=0.749$.

We considered 47 RRab and 40 RRc stars for an average distance of $9.48\pm0.57$ kpc.

\subsection{From the SX Phe P-L relation}

Yet another method towards the distance estimation is via the P-L relation for SX Phe stars. The presence of one SX Phe in M14, V177, enables this approach. The amplitude modulations in the light curve of V177 are clear in Fig. \ref{News} and suggest the presence of more than one mode. Our analysis identified two periods, P0=0.068984 d y P1=0.053607 d, for a ratio P1/P0=0.777, indicating that the modes correspond to  the radial fundamental and first overtone. The data fitted by a double-mode model with these periods are shown in Fig \ref{Doublemode}.

We have then employed the well known SX Phe P-L calibrations of \citet{Poretti2008}, \citet{Arellano2011} and \citet{CohenSara2012}.
The two period modes inserted in the three calibrations lead to an average distance of $d=8.6 \pm 0.3$ kpc.

\subsection{From the RR lyrae light curve Fourier decomposition}
\label{FourDist}

Fourier decomposition of the light curve leads to the calculation of the absolute magnitude via the calibrations of \citet{Kovacs2001} for RRab and of \citet{Kovacs1998} and the zero point calibration of \citet{Arellano2010} for the RRc stars. We found 9.15$\pm$0.99 kpc and  9.35$\pm$0.53 kpc respectively. See $\S$ \ref{secFOURIER} for details.

\subsection{From the models fitting in the CMD}
\label{CMDDist}

Positioning theoretical models of isochrones and ZAHB on the reddening-free CMD implies an assumption of the distance. Varying the distance until both loci are found to best represent the observations enables an estimation of the distance. We concluded that a distance of 9.3$\pm$0.1 kpc produces the best matching between observations and models (see CMD in Fig \ref{CMD}).

Table \ref{tabdistance} summarises our distance estimations to M14.

\begin{table}
\begin{center}
\caption{Distance to M14 from several methods} 
\label{tabdistance}
\begin{tabular}{lll}
\hline
Method&$d$ (kpc)&N\\
\hline
RRLs P-L ($I$) & 9.16$\pm$0.55&87\\
SX Phe P-L & 8.6$^{*}$&1\\
RRab Fourier decomposition & 9.61$\pm$0.67&24\\
RRc Fourier decomposition  & 9.39$\pm$0.48&36\\
ZAHB positioning  & 9.3$\pm$0.1 \\
\hline
Average&9.36$\pm$0.16& \\
\hline
\end{tabular}

\center{\quad $*$ Not included in average}
\end{center}
\end{table}

\section{Star membership using {\it Gaia}-eDR3}
\label{gaia}

Images of globular clusters are naturally contaminated by field stars that do not belong to the cluster. Presently the availability of the results of the Gaia project enables confident approaches to the discrimination between members and non-members. Before displaying and discussing the CMD of M14 we made use of the technique developed by \citet{Bustos2019}  to identify the likely cluster members. The method consists of two stages: the first stage is based on a clustering algorithm \citep{Zhang1996} employed to detect groups of stars in a four-dimensional space of physical parameters -projected positions and proper motions-; and the second stage,, which is an analysis of the projected spatial distribution of stars with different proper motions, enables the detection of members in the outskirts of the cluster, 
and of those with large proper motions dispersion typical in the central region. Finally the CMD in the Gaia photometric system is plotted in order to confirm that it is consistent with a globular cluster.

The analysis was carried for a 20 arcmin radius field centered in the cluster. The  field includes 42132 $Gaia$-eDR3 sources, although only 30392 have proper motions, and 10361 of these stars were found to be likely members. Our photometry includes 6767 of these member stars, on which we shall sustain the CMD discussion of x 9.
Fig.\ref{VPD} illustrates the resulting Vector-Point Diagram (VPD) showing the spatial motion of the cluster relative to the field stars, and the CMD in the Gaia photometric system.

\section{The colour-magnitude diagram of M14}
\label{secCMD}

 A Colour-Magnitude diagram is an important observational tool in which one can distinguish cluster star members from field stars in the FoV of the images, to appreciate the stellar distributions of very old low-mass stars (< 1 $M_{\odot}$) in their different evolutionary stages, and confront them with theoretical predictions. One can also consider the variable star distributions, in particular that of the two mode RR Lyrae stars, RRab and RRc, on the HB. In Fig. \ref{CMD} we display the CMD that emerged from our observations. In the top left panel the observational CMD, $(V-I)$ vs $V$ plane, is shown. Likely members of the cluster are plotted with black dots whereas likely field stars are represented by small light blue dots. All variables, previously known and newly discovered are plotted with colours, coded as indicated in the caption. The top right panel shows the reddening free version of the CMD, $(V-I)_o$ vs $V_o$, for which we have adopted the differential reddening map provided by CP13. In order to apply the individual reddenings of CP13, we had to match each star in our field with those in CP13's field of M14. The equatorial coordinates were used for this purpose, however since their reference image and our own come from different astrometrical solutions, small coordinate differences may be expected. We accepted matches when they were better that 2 arc seconds, i.e. from the 6767 members in the left panel, we safely deredden 5937 stars. Therefore, the upper right panel of Fig. \ref{CMD} is built on likely cluster members duely corrected from the differential reddening effects. In order to match the observations with the theoretical models, an average overall reddening of $E(B-V)=0.59$ or $E(V-I)=1.259 \times E(B-V)=0.74$ was adopted. Other larger values found in the literature ($\S \ref{reddening}$) move the stars too far to the blue in the CMD. All presently known variables in our study (Table \ref{tab:variables}) are shown with colours according to their type. We have shifted vertically, until reach a reasonable fit to the data, the isochrones and ZAHB models described in the caption. We found that a good agreement with the data is obtained for a distance of 9.3$\pm$0.1 kpc. Finally, in the bottom panel, an expansion of the HB region is shown. Here we include the empirical determination of the FORE at $(V-I)_o$ =0.45-46 as determined from homogeneous data from several clusters by \citet{Arellano2015,Arellano2016}. It is clear that some RRab stars are sitting to the blue of the FORE, i.e. in the "either-or" or bimodal region shared by the fundamental and first overtone pulsators RRab and RRc. Thus, in spite of its Oo-int nature, in M14 the two modes are not neatly segregated as is the case in all OoII and only in some OoI globular clusters. Whether this is a consequence of more than one stellar generation being present in the cluster, it is something we can neither rule out nor proof. The [Fe/H]$_{\rm ZW}$-$\cal L$ plane is displayed in Fig. \ref{MetHB} for a family of clusters of assorted Oosterhoff types, with an indication of those clusters where the two modes, i.e. the first overtone or RRc and the fundamental mode or RRab stars, are segregated, and those clusters where the bimodal region is shared by the two modes. At this point we should call attention to the RRc stars sitting to the red of the FORE, i.e. outside the first overtone instability strip. This anomalous position is mild for some stars (V90, V107, V135 and V175) and gross for others (V126, V131). It is pertinent to mention that most of them were similarly found by CP18 in an inappropriate position in the $(B-V)-V$ CMD (and others not considered here; V121 and V127, see their figure 4). All these stars are within $\sim$ 0.5-1 arcmin from the cluster center. Thus, it is very likely that their odd position is due to colour contamination by an undetected close neighbour. An inspection of the $Gaia$-eDR3 sources revealed that most of them have a neighbouring source very near or even within their PSF. These stars shall be further addressed in Appendix B.
 
As already noted by CP13, the age estimation of M14 has not been included in major studies of cluster ages of the last 20 years. These authors however attempted an age determination for M14 by the vertical method, or the magnitude difference between the turn off point and the HB, and a comparison of the fiducial RGB on the $(B-V)-V$ plane with that of \citet{Sandquist1996} for M5. They came to the conclusion that M14 and M5 are of similar age or, if anything, M14 may be a bit older than M5, errors permitting, by as much as 0.3 Gys. The absolute age of M5 has been estimated by \citet{Dotter2010} using relative ages from isochrone fitting, and by \citet{VandenBerg2013} using an improved calibration of the vertical method; these authors find 12.25±0.75 Gyr and 11.50±0.25 Gyr respectively. Given the above results, it seems reasonable to assume the age of M14 as 12-12.5 Gyrs. We overlaid isochrons of 12.0 and 12.5 Gyrs to our CMD which indeed represent satisfactorily our observations.

Assuming the above age and a metallicity of  Z = 0.001 ($\sim$ [Fe/H] = -1.3), we used the Eggleton code \citep{Pols1997, Pols1998, KPS1997} and a modified mass loss Reimers law  \citep{SC2005} to produce ZAHBs models for different He core masses. In Fig. \ref{CMD}, bottom panel, we illustrate the cases for core masses 0.48 (lilac), 0.49 (red) and 0.50 (green) $M_{\odot}$, with a total masses range of 0.52-0.63 $M_{\odot}$ which cover the whole HB.  Notice how for larger core masses the ZAHB naturally becomes more luminous. For comparison, we included the ZAHB of the \citet{Vandenberg2014} models (segmented blue line) with similar input parameters. The agreement is very good.

It is clear from the CMD of M14 that the HB, RGB and TO regions all look rather wide. Although part is certainly due to the intrinsic uncertainties of our photometry, real He-abundance variations due to the presence of more than one stellar generation can play a prominent role. The presence of more than one generation of stars in GCs is by now a well known fact \citep[e.g.][]{Bedin2004,Piotto2005,Piotto2007}. It has been shown \citep{Milone2017} that at least two generations of stars, G1 and G2, coexist in globular clusters, which in fact have been identified in 57 CGs
using adequate colour-colour and colour-magnitude combinations and accurate {\it Hubble Space Telescope} photometry. The He mass fraction in the 2G has been found to be enhanced by between 0.01 to 0.10
and that the He-richer the 2G is, the more extended appears the HB \citep{Milone2018}.

That helium shell content could be responsible for the width of the HB, we have confirmed by overlaying HB versions of different Y values from the models of \citet{Vandenberg2014}. A range of Y of 0.23-0.31 is sufficient to explain the observed width of the HB.

While the structure of the HB is defined by multiple generations of different He abundances, we highlight that having a larger core mass, as we have used above, produce the same effect of increasing the luminosity, since in both cases the density and pressure in the hydrogen-burning shell, which still is the stronger contributor to the total HB luminosity rises, albeit for different reasons (larger mean molecular weight and higher gravity, respectively). We can therefore not distinguish between these two possibilities. However, for the interpretation of the observations presented here, there is no fundamental difference.

Our models also indicate that a 0.84-0.85 $M_{\odot}$ progenitor star in the main sequence reaches the tip of the RGB in about 11.9-12.5 Gyrs, thus, by the time these stars reach the ZAHB, they have lost $\sim$ 0.22-0.32 $M_{\odot}$ during the He-flash events at the RGB. 
One may wonder, how much that depends on the mass loss prescription chosen. 
In fact, \citet{SC2005} actually calibrated the modified Reimers mass loss prescription, using the same evolution code and parametrization of RGB stars
as we do here, to give the canonical helium core masses of 0.48-0.50 $M_{\odot}$ of HB stars, and that so for very different metallicities and for globular clusters by virtue 
of the added terms. Hence, regardless of the actual mass loss prescription, it is that calibration of the total mass lost on the RGB which matters here.
At the same time, \citet{SC2005} suggested that some
stars may have up to 25\% less mass loss than those with the thinnest
shells (those marking the blue end of the HB), which could give rise to
slightly larger He core masses.
Let us now explore the evolution of these resulting low mass stars as they evolve off the ZAHB, and compare our resulting evolutionary tracks with the locus of the type II Cepheids present in the cluster.

\section{post-ZAHB evolutionary models }
\label{sect:HRD}

\begin{figure}
\begin{center}
\includegraphics[width=8.2cm]{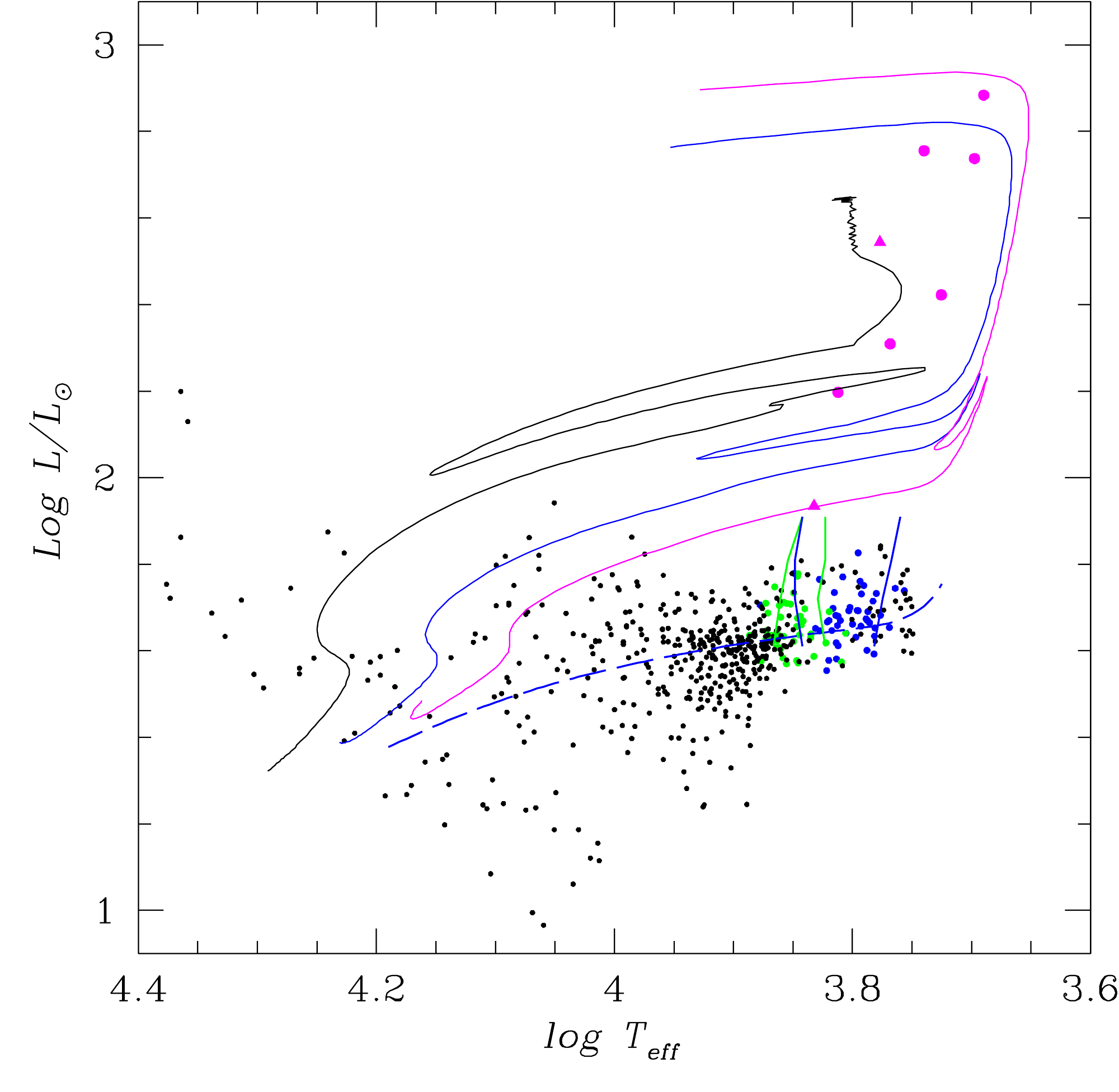}
\caption{Post He-flash models for a He core mass of 0.48 $M_{\odot}$. Black, blue and lilac lines corresponds to models with total masses of 0.52, 0.535 and 0.55 $M_{\odot}$, respectively. The ZAHB segmented blue line is from \citet{Vandenberg2014} as in the Fig. \ref{CMD}. Blue and green dots represent RRab and RRc star respectively. The instability strip border lines are the theoretical predictions of \citet{Bono1994} for the  fundamental mode (blue) and the first overtone (green). See $\S$ \ref{sect:HRD} for a discussion.}
    \label{HR}
\end{center}
\end{figure}

M14 has a good number of type II Cepheid stars, or BL Her (CWB) and W Vir (CWA) stars. Thus, this cluster offers a good chance to compare our observations with evolutionary post ZAHB models. In order to better visualise a comparison of the theoretical predictions and the observations, we choose to convert our CMD into the HRD of Fig. \ref{HR}, by transforming the observed $(V-I)-V$ into log $T_{\rm eff}$-log$(L/{\rm L_{\odot}})$. In order to do this, we adopted the $(V-I)_o$-log $T_{\rm eff}$ and $BC$-log $T_{\rm eff}$ calibrations of \citet{VdB2003}. Notice the larger scatter in the stellar distribution on the HB as we move to the blue, since these stars correspond to the bottom of the blue tail of the HB in the CMD, and therefore are mostly faint stars that may or may not truly belong to the HB. Fig. \ref{HR} shows the evolutionary tracks for stars with a He core of 0.48 $M_{\odot}$; the black, blue and lilac tracks correspond to total masses of 0.52, 0.535, and 0.55 $M_{\odot}$ respectively, i.e. stars with a very thin envelope of 0.04-0.07 $M_{\odot}$. We preferred the 0.48 $M_{\odot}$ core mass as opposed to 0.49 and 0.50 $M_{\odot}$ since the latter naturally produce tracks that are too luminous relative to the adopted ZAHB. 

In Fig. \ref{HR} the distribution of RRab (blue dots) and RRc stars (green dots) is also displayed, along with the theoretical borders of the instability strip for the fundamental and first overtones modes \citep{Bono1994}. In this process, we have employed $E(B-V)=0.59$ as discussed previously, and notice that other values, such as 0.61, would corrupt the matching between the theoretical and the observational distribution of RR Lyrae stars and the instability strip. Our masses are in good agreement with the predictions of \citet{Bono1997} of 0.52-0.59 $M_{\odot}$ total masses for type II Cepheids. Thinner envelope models are  hotter and display longer blue loops crossing the IS above the HB, in the region of BL Her and W Vir stars. The same is true for a He core of 0.49 and 0.50 $M_{\odot}$, with a very thin envelope of 0.04 $M_{\odot}$, however, as mentioned before these models turn out to be a bit too luminous.

We conclude that our post He flash evolutionary models in the total mass range of 0.52-0.63 $M_{\odot}$, and He core range of 0.48-0.50 $M_{\odot}$, represent very well both the stretch of the HB in $T_{\rm eff}$, as well as in luminosity. The former given by some spread in the shell mass left behind by the RGB mass losses, and the latter given by the rising luminosity of these HB stars during central He burning. It seems even plausible to interpret the presence of numerous type II Cepheids of M14 as products of post-HB evolution, driven by the complex processes involving the burning of the very thin, low mass hydrogen and helium shells of these stars and their minuscule envelopes, as they ascend on their AGB.

\section*{ACKNOWLEDGEMENTS}
MAY is grateful to CONACyT (Mexico) for a Ph.D. scholarship. This project was supported by the program PAPIIT, UNAM, through grant IG100620. 
We thank the staff of SPM, Baja California and IAO, Hanle and CREST, Hosakote, for making these observations possible. The facilities at IAO and CREST are operated by the Indian Institute of Astrophysics, Bangalore.
 We have made extensive use of the
SIMBAD and ADS services, for which we are thankful.

\section*{DATA AVAILABILITY}
The data underlying this article shall be available in an electronic
form in the Centre de Donnés astronomiques de Strasbourg data
base (CDS), and can also be shared on request to the corresponding
author.

\bibliographystyle{mnras}
\bibliography{M14} 

\appendix
\label{appendix}

\section{Comparison with CP18 photometry and peculiar stars}
\label{AppA}

\subsection{Comparison with CP18 photometry}
\label{SecA1}

In this section, we compare our $V$ photometry and that of CP18. Fig. \ref{comp} shows the magnitude differences  $V_{\rm Yep}-V_{\rm CP18}$ as a function of the distance (in arcsec) to cluster center. Note the larger difference towards the cluster center, as expected since the central region is highly contaminated by undetected faint neighbours, affecting both photometric sets. Considering the different approaches to the transformation to the standard system in both studies, the magnitude level difference compares rather well. The mean difference value is -0.037, in the sense that our photometry is in average a bit brighter than CP18's.

In Fig. \ref{comp} we have labeled some stars with particularly large differences. Stars V61, V75, and V129 show, however, neat well defined light curves in our data and we find no obvious contamination. For stars V73, V114, and V172, we find clarifying reasons for the large difference as discussed in the following section.

\subsection{Comments on peculiar stars}
\label{SecA2}

\textbf{V55}, \textbf{V58}, and \textbf{V153}. The light curves of these RRc variables show prominent amplitude modulations. We have searched for evidence of double mode oscillations but found no convincing results. Hence, there is the possibility that these stars are Blazhko variables.

\begin{figure}
\begin{center}
\includegraphics[width=8.4cm]{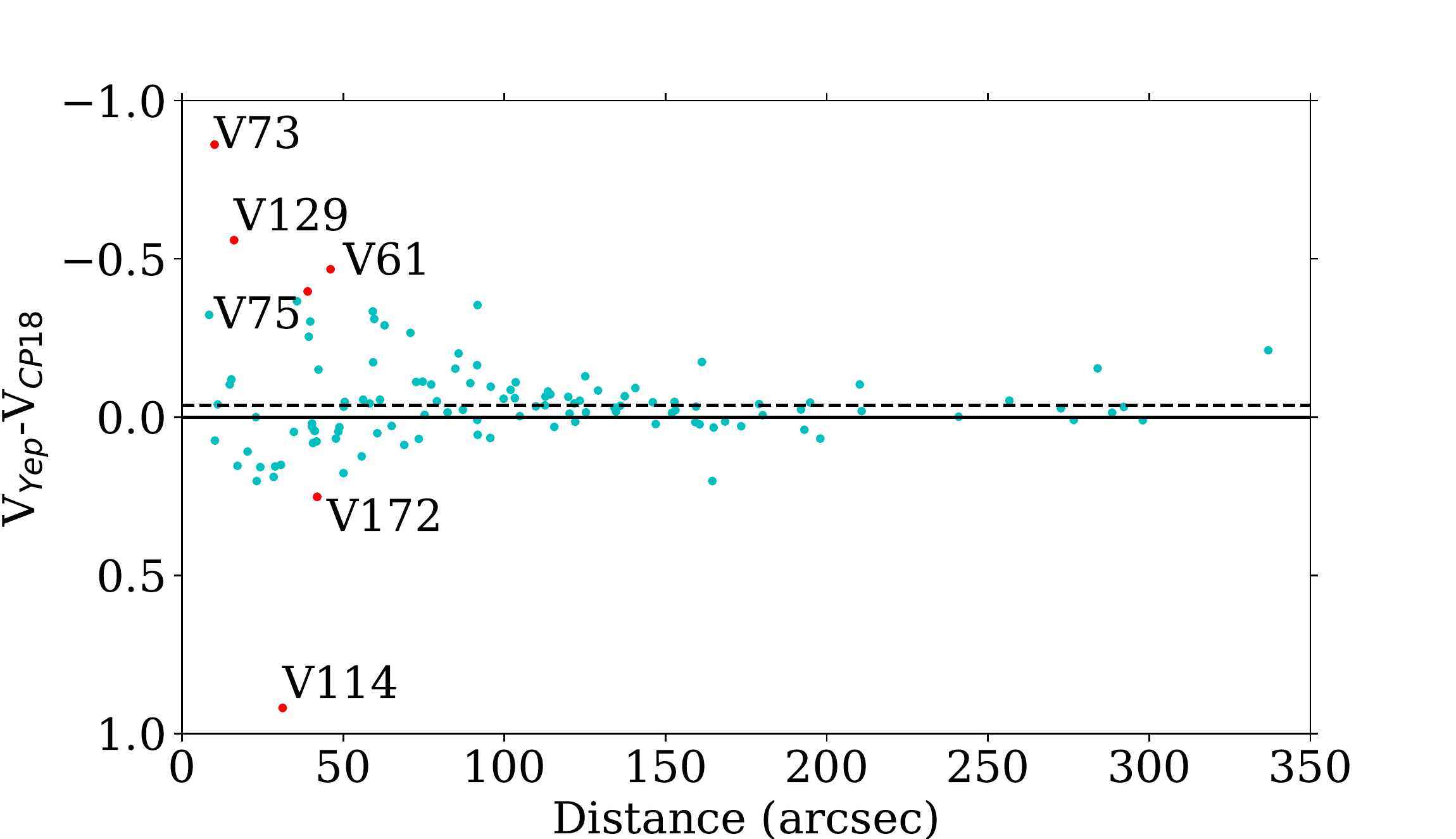}
\caption{Mean difference between our $V_{\rm Yep}$ magnitude and CP18's,  $V_{\rm CP18}$,  as function of the distance to the cluster center  (in arcsec). $V_{\rm Yep}$ is a little brighter than that of CP18 by  $-0.037$ mag (dashed line). Larger difference towards the cluster center are due to the higher contamination of undetected faint neighbours affecting both photometric sets. Labeled stars are discussed in  $\S$ \ref{SecA2}.}
    \label{comp}
\end{center}
\end{figure}

\textbf{V73}. This is an SR variable star located in the crowded cluster central region, and therefore likely contaminated by neighboring stars which may explain the difference between our and CP18's $V$ mean magnitudes. However, the large difference could also be due to the fact that CP18 only observed a short variability during the light rise and also we did not observe the complete curve.

\textbf{V90}, \textbf{V107}, \textbf{V126}, \textbf{V131} and \textbf{V135}. Although from their $Gaia$-eDR3 proper motion analysis these stars were considered cluster members (except V131 without data), in the CMD these RRc stars are to the red of the FORE. i.e. outside the instability strip of the first overtone. V126 and V131 are also well above the HB.
Similar outstanding positions were reported by CP18 for the majority of them. Their odd position is likely due to light contamination from a close undetected neighbour.  From the $Gaia$-eDR3 source catalogue it is clear that all these stars are within 1 arcmin from the cluster center and  have at least one close neighbour star even within their PSF, therefore, it is very likely that their observed colour is spurious.

\textbf{V114}. The mean \emph{VI} magnitudes of our light curves, $V$=17.299 and $I$=16.170 show a star that is about one magnitude fainter than the reported by CP18. We should emphasize, however, that our light curve has less dispersion, and with the above magnitude and colour the star lies in the expected region of the HB. According to the $Gaia$-eDR3 database,  V114 is among 4 other point sources, which are likely responsible for the light contamination and the alteration of the mean values of  CP18.

\textbf{V119} and \textbf{V169}. The light curves of these two variables are very scattered and a close examination of the $Gaia$ sources on our reference images reveals the presence of at least two stars within their PSF, hence there is a clear contamination of non-resolved close neighbour. We did not consider these stars in this work.

\textbf{V120}. A small refinement to the coordinates given in the CVSGC confirms the variable to be a bright star of the SR type. The  $Gaia$-eDR3 reports only one source in the region, corresponding to the bright star.

\textbf{V155}. CP18 were unable to detect variations in this star. Our data suggest mild variations that can in fact be phased with a period of 36.54 d, suggesting it to be a SR variable, properly placed at the CMD. Further data are needed to confirm this conclusion.

\textbf{V172}. The light curve of this RRab star in our data shows a much larger amplitude than that reported by CP18. While this difference could possibly be due to Blazhko-like modulations, it should be noted that the mean magnitude difference is large, +0.25 mag, (Fig \ref{comp}), therefore, being the star in the central regions of the cluster, we cannot rule out the possibility of an undetected but substantial light contamination.

\textbf{V175}. The star is found to be variable with a period of 0.299130 d and we classified it as RRc. It lies however very near the RRab star V43, for which we found a period of 0.521738 d. Despite their closeness, we have been able to verify, by blinking the residual images, that both the stars are authentic variables. Although the amplitude of V175 suggests modulations, we have refrained from a further analysis given the likely light contamination from V43.

\bsp	
\label{lastpage}
\end{document}